\documentclass[11pt,preprint,nofootinbib,aps,twoside,showpacs,showkeys]{revtex4}

\usepackage{graphicx}
\usepackage{hyperref}
\usepackage{pdfsync}

\usepackage{amssymb}

\def\Tr{\mbox{Tr}\,}

\newcommand{\R} {\mbox{Re}\,}

\newcommand{\be}{\begin{equation}}
\newcommand{\ee}{\end{equation}}
\newcommand{\bea}{\begin{eqnarray}}
\newcommand{\eea}{\end{eqnarray}}
\newcommand{\eu}{{\rm e}}
\newcommand{\ii}{{\rm i}}
\newcommand{\de}{{\displaystyle\rm\mathstrut d}}
\newcommand{\Ord}{{\rm O}}
\newcommand{\mod}{{\rm \: mod \:}}
\newcommand{\Pf}{{\rm Pf}}

\def\Xint#1{\mathchoice
   {\XXint\displaystyle\textstyle{#1}}%
   {\XXint\textstyle\scriptstyle{#1}}%
   {\XXint\scriptstyle\scriptscriptstyle{#1}}%
   {\XXint\scriptscriptstyle\scriptscriptstyle{#1}}%
   \!\int}
\def\XXint#1#2#3{{\setbox0=\hbox{$#1{#2#3}{\int}$}
     \vcenter{\hbox{$#2#3$}}\kern-.5\wd0}}

\def\dashint{\Xint-}

\begin{document}

\today

\title [EFP for the XY Spin Chain in a Magnetic Field]
{Asymptotics of Toeplitz Determinants and the Emptiness Formation
Probability for the XY Spin Chain}

\author{Fabio Franchini and Alexander G. Abanov
\footnote[3]{E-mail: Fabio.Franchini@sunysb.edu;
 Alexandre.Abanov@sunysb.edu} }

\affiliation{Physics \& Astronomy Department,
\\ Stony Brook University,
\\ Stony Brook, New York 11794-3800}

\begin{abstract}

We study an asymptotic behavior of a special correlator known as
the Emptiness Formation Probability (EFP) for the one-dimensional
anisotropic XY spin-1/2 chain in a transverse magnetic field.
This correlator is essentially the probability of formation of a
ferromagnetic string of length $n$ in the antiferromagnetic ground
state of the chain and plays an important role in the theory of
integrable models.
For the XY Spin Chain, the correlator can be expressed as the
determinant of a Toeplitz matrix and its asymptotical behaviors for
$n\to \infty$ throughout the phase diagram are obtained using known
theorems and conjectures on Toeplitz determinants.
We find that the decay is exponential everywhere in the phase diagram
of the XY model except on the critical lines, i.e. where the spectrum
is gapless.
In these cases, a power-law prefactor with a universal exponent arises
in addition to an exponential or Gaussian decay.
The latter Gaussian behavior holds on the critical line corresponding
to the isotropic XY model, while at the critical value of the magnetic
field the EFP decays exponentially.
At small anisotropy one has a crossover from the Gaussian to the
exponential behavior.
We study this crossover using the bosonization approach.
\end{abstract}



\maketitle

\maketitle

\pagenumbering{Roman}

\tableofcontents

\newpage

$\qquad$

\newpage

\pagenumbering{arabic}

%%%%%%%%%%%%%%%%%%%%%%%%%%%%%%%%%%%%%%%%
\section{Introduction}
%%%%%%%%%%%%%%%%%%%%%%%%%%%%%%%%%%%%%%%%

Although the Bethe Ansatz \cite{bethe,yangyang,korepin93} gives us
important information about the ground state and thermodynamics of
quantum integrable one-dimensional models, the calculation of
correlation functions for these models is still an open problem.
Considerable progress, however, has been made recently in this field
(see \cite{korepin93,jimbo} and references therein).

Two main approaches express correlation functions in integrable models
in terms of determinants of Fredholm operators \cite{korepin93,korepin94}
or as multiple integrals \cite{jimbo}.
These expressions are exact but are very complex. 

It was argued \cite{korepin93} that the simplest of the correlators in
some integrable models is the so-called "Emptiness Formation Probability"
(EFP).
For an XYZ spin chain in a magnetic field which is defined as
\be
 \label{XYZh}
   H = \sum_{i=1}^N \left[ J_x \sigma_i^x \sigma_{i+1}^x +
   J_y \sigma_i^y \sigma_{i+1}^y + J_z \sigma_i^z \sigma_{i+1}^z \right]
   - h \sum_i \sigma_i^z,
\ee
the Emptiness Formation Probability represents the "Probability of
Formation of Ferromagnetic Strings"
\be
   P(n) \equiv {1 \over Z} \Tr \left\{ \eu^{- \frac{H}{T}}
   \prod_{j=1}^n {1 - \sigma_i^z \over 2} \right\},
   \label{EFPNonZeroTDef}
\ee
where $Z \equiv \Tr \left\{ \eu^{- H/T} \right\}$.
At zero temperature $T=0$ it becomes
\be
   P(n) \equiv
   \langle 0 | \prod_{i=1}^n {1 - \sigma_i^z \over 2} | 0 \rangle ,
   \label{EFPDef}
\ee
that is, the probability that $n$ consecutive spin sites are all found
aligned downward in the ground state $| 0 \rangle$.

It is conceivable that the study of this simple correlator will bring
insights helpful to carry on the investigation of other correlators.
But despite the fact that the EFP is the simplest of the correlators and
despite considerable efforts devoted over the years to its study,
there is still no easy recipe for its calculation (see, for
instance, \cite{korepin93} -- \cite{KLNS-2002}).
However, some progress has been achieved in finding an asymptotic
behavior of $P(n)$ at large distances $n \rightarrow \infty$.

For the XXZ spin chain in zero magnetic field ($J_{x}=J_{y}$, $h=0$ in
(\ref{XYZh})), the EFP is found to be Gaussian $P(n)\sim e^{-\alpha n^{2}}$
as $n \to \infty$ in the critical regime $|J_{z}|<|J_{x}|$ at zero
temperature and exponential $e^{-\beta n}$ at finite temperature
(\cite{KMST_gen-2002},\cite{KLNS-2002}).

A qualitative argument in favor of Gaussian decay was given in
Ref. \cite{abanovkor} within a field theory approach.
It was argued there that the asymptotics of the EFP are defined by the
action of an optimal fluctuation (instanton) corresponding to the EFP.
In the critical model, this fluctuation will have a form of a
``$n\times n$'' droplet in space-time with the area $A \sim n^{2}$ and
the corresponding action $S \approx \alpha n^{2}$ which gives the decay
$P(n)\sim e^{-\alpha n^{2}}$.
Similarly, at finite temperature the droplet becomes rectangular (one
dimension $n$ is replaced by an inverse temperature $T^{-1}$) and the
action cost is proportional to $n$, giving $P(n) \sim e^{-\beta n}$.
This argument is based on the criticality of the theory\footnote{More
precisely, on the assumption that temporal and spatial dimensions of an
instanton scale similarly.} and it is interesting to consider whether
it could be extended to a non-critical theory.
A na\"ive extension of the argument would give the optimal
fluctuation with space-time dimensions $n \times \xi$ where $\xi$ is a
typical correlation length (in time) of the theory.
This would result in $P(n) \sim e^{-\beta n}$ for
non-critical theories, similarly to the case of finite temperature in
critical regime.
The rate of decay $\beta$ would be proportional to the correlation
length of the theory.

In this paper we examine the relation between the asymptotic behavior
of the EFP and criticality using the example of the Anisotropic XY spin-1/2
chain in a transverse magnetic field
\be
   H = \sum_{i=1}^N \left[
   \left( {1 + \gamma \over 2} \right) \sigma_i^x \sigma_{i+1}^x +
   \left( {1 - \gamma \over 2} \right) \sigma_i^y \sigma_{i+1}^y \right]
   - h \sum_{i=1}^N \sigma_i^z,
   \label{spinham}
\ee
where $\sigma_i^{\alpha}$, with $\alpha=x,y,z$, are the Pauli
matrices which describe spin operators on the $i$-th lattice site of
the spin chain and, for definiteness, we require periodic boundary
conditions: $\sigma_i^\alpha = \sigma_{i+N}^\alpha$ ($N >> 1$).
This model is, probably, the simplest model that has both critical
and non-critical regimes in its $\gamma-h$ phase diagram.

Using the mapping of the model (\ref{spinham}) to free one-dimensional
fermions (Jordan-Wigner transformation) one can express the EFP $P(n)$
of this system in terms of a determinant of a $n \times n$ Toeplitz
matrix.
The asymptotic behavior of these matrices can be found using known
theorems from the theory of Toeplitz determinants.
This approach and technique is exactly the one of Ref. \cite{mccoy},
where it was used to find the spin-spin correlation functions for the
model (\ref{spinham}).
This technique was also used in Ref. \cite{shiroishi} for the EFP in the
case of the Isotropic XY model (Eq. (\ref{spinham}) with $\gamma=0$).
In the latter work it was shown that the EFP decays in a Gaussian way
for the critical theory ($\gamma=0$, $-1 \le h \le 1$).
This case corresponds to one of the two critical lines in the
$\gamma-h$ phase diagram of the model (\ref{spinham}).
The other line is the critical magnetization line(s) ($h = \pm 1$).
In the rest of the two-dimensional $\gamma-h$ phase diagram, the model
is non-critical.

We obtain that the EFP is asymptotically exponential in most of the
phase diagram according to the na\"ive expectations and that it is
Gaussian only at $\gamma=0$ in agreement with previous works on XXZ spin
chains and Ref. \cite{shiroishi}.
However, on the critical lines $h =\pm 1$, in addition to the exponential
decay, a pre-exponential power-law factor arises, with a universal
exponent.
The power-law prefactor is present in the isotropic case as well, but
with a different exponent.

The paper is organized in the following way: we briefly review the phase diagram of the XY
Spin-1/2 Chain in Section \ref{XYModel}.
In Section \ref{EFPinXY} we explain how one can express the EFP as the
determinant of a Toeplitz matrix and review our results
so that readers who are not interested in derivations can skip the
next sections.
In Section \ref{expsection} we analyze the exponential decay of the EFP
for the non-critical and critical phases of the anisotropic XY Model.
In Section \ref{pre-ex} we derive in detail the asymptotic behaviors,
including the pre-exponential factors, of both non-critical and critical
parts of the phase diagram.
In Section \ref{GammaESec} we study a special line of the phase diagram
on which the ground state is known exactly and compare the explicit
results one can obtain using the exact ground state with the asymptotes of
the EFP we derived in the previous sections.
In Section \ref{gammazero} we report on the already known results for
the EFP of the isotropic XY model \cite{shiroishi}.
In Section \ref{Crossover} we make contact with Ref. \cite{abanovkor}
using a bosonization approach to discuss the crossover as a function of $n$ 
from the Gaussian
to the exponential behavior of EFP for the case of small anisotropy $\gamma$.
Finally, Section \ref{Conclusions} will summarize our results.
For the reader's convenience we collect some results on asymptotic
behavior of Toeplitz determinants which are extensively used in the
rest of the paper in the first appendix.
The second appendix presents the analysis of the finite temperature
behavior of the EFP, which gives an expected exponential decay.
The third appendix gives some mathematical details on the calculation
of the stationary action in the bosonization approach of Section 
\ref{Crossover}.

Some of the results presented in this paper have been announced in a
previous letter \cite{abanovfran} without details, which will be given
here.

%%%%%%%%%%%%%%%%%%%%%%%%%%%%%%%%%%%%%%%%
\section{The Anisotropic XY Model}
\label{XYModel}
%%%%%%%%%%%%%%%%%%%%%%%%%%%%%%%%%%%%%%%%

The XY spin model defined by (\ref{spinham}) has been solved in
\cite{LSM-1961} in the case of zero magnetic field and in \cite{mccoy}
in the presence of a magnetic field.
We follow the standard prescription \cite{LSM-1961} and reformulate the
Hamiltonian (\ref{spinham}) in terms of spinless fermions $\psi_i$ by
means of a Jordan-Wigner transformation:
\bea
   \sigma_j^+ & = &
   \psi_j^\dagger \eu^{\ii \pi \sum_{k<i} \psi_k^\dagger \psi_k} =
   \psi_j^\dagger \prod_{k<j}
   \left( 2 \psi_k^\dagger \psi_k - 1 \right), \\
   \sigma_j^z & = & 2 \psi_j^\dagger \psi_j - 1,
\eea
where, as usual, $\sigma^\pm = (\sigma^x \pm i \sigma^y)/2$:
\be
   H = \sum_{i=1}^N \left( \psi_i^\dagger \psi_{i+1} +
   \psi_{i+1}^\dagger \psi_i +
   \gamma \; \psi_i^\dagger \psi_{i+1}^\dagger +
   \gamma \; \psi_{i+1} \psi_i -
   2 h \; \psi_i^\dagger \psi_i \right) .
   \label{realfermionH}
\ee
In Fourier components $\psi_j = \sum_q \psi_q \eu^{\ii q j}$, we have:
\be
   H = \sum_q \left[
   2 \left( \cos \! q - h \right) \psi_q^\dagger \psi_q
   + \ii \gamma \sin \! q \: \psi_q^\dagger \psi_{-q}^\dagger
   - \ii \gamma \sin \! q \: \psi_{-q} \psi_q \right].
   \label{spinlessham}
\ee
The Bogoliubov transformation
\be
   \chi_{q} = \cos \! {\vartheta_q \over 2} \: \psi_{q}
   + \ii \sin \! {\vartheta_q \over 2} \: \psi_{-q}^\dagger
   \label{bogtrans}
\ee
with ``rotation angle'' $\vartheta_q$
\be
   e^{i\vartheta_q}  = \frac{1}{\varepsilon_q}(\cos q-h+i\gamma\sin q),
   \label{rotangle}
\ee
brings the Hamiltonian (\ref{spinlessham}) to the diagonal form
$\sum_q \varepsilon_q \chi_q^\dagger\chi_q$ with the quasiparticle
spectrum
\be
   \varepsilon_q = \sqrt{ \left( \cos q - h \right)^2
   + \gamma^2  \sin^2 q}.
   \label{spectrum}
\ee
We recognize from (\ref{spectrum}) that the theory is critical, i.e.
gapless, for $h=\pm 1$ or for $\gamma = 0$ and $|h| < 1$.

\begin{figure}
  \includegraphics[width=\columnwidth]{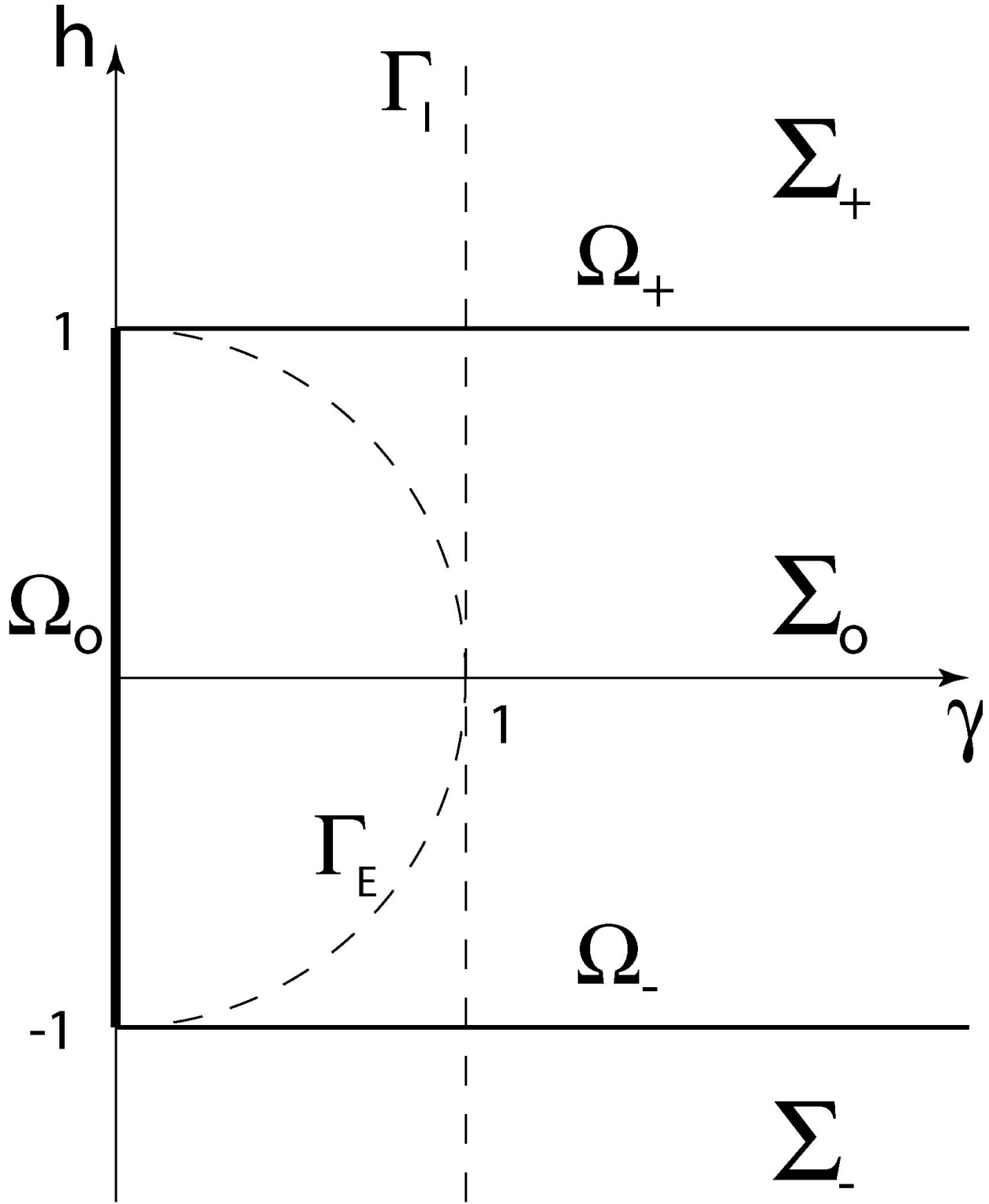}
\caption{Phase diagram of the XY Model (only the part $\gamma \ge 0$ is shown).
The theory is critical for $h = \pm 1$ ($\Omega_\pm$) and for $\gamma =
0$ and $|h| < 1$ ($\Omega_0$). The line $\Gamma_I$ represents the Ising
Model in transverse field. On the line $\Gamma_E$ the ground state of
the theory is a product of single spin states.}
   \label{phasediagram}
\end{figure}

In Fig.~\ref{phasediagram} we show the phase diagram of the XY model,
which has obvious symmetries $\gamma \to -\gamma$ and $h \to -h$.
However, the latter is not a symmetry of the EFP.
Therefore, we show only the part of the diagram corresponding to
$\gamma \ge 0$.
The phase diagram has both critical and non-critical regimes.
Three critical lines $\Omega_0$ (Isotropic XY model: $\gamma=0$, $|h|<1$)
and $\Omega_\pm$ (critical magnetic field: $h = \pm 1$) divide the
phase diagram into three non-critical domains, $\Sigma_-$, $\Sigma_0$,
and $\Sigma_+$ ($h < -1$, $-1 < h < 1$, and $h > 1$ respectively).
Fig.~\ref{phasediagram} also shows the line $\gamma=1$ ($\Gamma_I$)
corresponding to the Ising model in transverse magnetic field and the
line $\gamma^2 + h^2 = 1$ ($\Gamma_E$) on which the wave function of the
ground state is factorized into a product of single spin states
\cite{shrock} (we will consider this line in detail in Section
\ref{GammaESec}).

The fermionic correlators are easy to obtain from (\ref{spinlessham}).
In the thermodynamic limit they read \cite{LSM-1961,mccoy}
\bea
   F_{jk} &\equiv& \ii\langle \psi_j \psi_k \rangle
   = - \ii\langle \psi_j^\dagger \psi_k^\dagger \rangle
   = \int_0^{2 \pi} {\de q \over 2\pi}\; \frac{1}{2}\sin\vartheta_q
   \eu^{\ii q (j-k)},
  \label{F}
\\
   G_{jk} &\equiv& \langle \psi_j \psi_k^\dagger \rangle
   =   \int_0^{2 \pi} {\de q \over 2 \pi}\; \frac{1+\cos\vartheta_q}{2}
   \eu^{\ii q (j-k)} .
 \label{G}
\eea

%%%%%%%%%%%%%%%%%%%%%%%%%%%%%%%%%%%%%%%%
\section{Emptiness Formation Probability in the XY model}
\label{EFPinXY}
%%%%%%%%%%%%%%%%%%%%%%%%%%%%%%%%%%%%%%%%

We are mostly interested in the value of the correlator
(\ref{EFPNonZeroTDef}) at zero temperature (the non-zero temperature
case is deferred to the second appendix).
In terms of spinless fermions, one can express the EFP (\ref{EFPDef})
as the expectation value over the ground state of the theory
\cite{shiroishi}
\be
   P(n) = \langle \prod_{j=1}^n \psi_j \psi_j^\dagger \rangle.
   \label{expect}
\ee
This expression projects the ground state on a configuration without
particles on a string of length $n$ and gives the meaning to the name
``Emptiness Formation Probability''.

Let us introduce the $2n \times 2n$ skew-symmetric matrix ${\bf M}$ of
correlation functions
\be
   {\bf M} = \pmatrix{  {-\ii\bf F} &
                        {\bf G}  \cr
                        - {\bf G}  &
                        \ii{\bf F} \cr
                     },
\ee
where ${\bf F}$ and ${\bf G}$ are $n \times n$ matrices with matrix
elements given by $F_{jk}$ and $G_{jk}$ from (\ref{F},\ref{G})
respectively.
Then, using Wick's theorem on the r.h.s of (\ref{expect}), we obtain
\be
    P(n) =  \Pf( {\bf M} ).
\ee
The Pfaffian \cite{mehta2} is defined as
\be
   \Pf( {\bf M} ) \equiv
   \sum_P (-1)^P M_{p_1p_2} M_{p_3p_4} \ldots M_{p_{2n-1}p_{2n}},
\ee
where $P=\{p_1,p_2,\ldots, p_{2n}\}$ is a permutation of
$\{1,2,\ldots,2n\}$, the sum is performed over all possible
permutations, and $(-1)^P$ is the parity of the permutation.
Using one of the properties of the Pfaffian we have
\be
    P(n) =  \Pf( {\bf M} ) = \sqrt{\det( {\bf M} )}.
\ee

We perform a unitary transformation
\be
   {\bf M'} = {\bf U M U}^{\dagger}
   = \pmatrix { 0 & {\bf S_n} \cr -{\bf S_n}^\dagger & 0 \cr}, \qquad
   {\bf U} = {1 \over \sqrt{2}} \pmatrix { {\bf I} & {\bf -I} \cr
                                           {\bf I} & {\bf I} \cr},
\ee
where ${\bf I}$ is a unit $n \times n$ matrix and ${\bf S_n}={\bf
G}+\ii{\bf F}$ and ${\bf S_n}^\dagger={\bf G}-\ii{\bf F}$.
This allows us to calculate the determinant of ${\bf M}$ as
\be
   \det ( {\bf M} ) = \det ( {\bf M'} ) =
   \det ( {\bf S_n} ) \cdot \det ( {\bf S_n}^\dagger )
   = \left|\det ( {\bf S_n} ) \right|^2.
   \label{detM}
\ee
The matrix ${\bf S_n}$ is a $n \times n$ Toeplitz matrix (i.e. its
matrix elements depend only on the difference of row and column
indices \cite{basor}).
The generating function $\sigma(q)$ of a Toeplitz matrix is defined by
\be
    ({\bf S_n})_{jk}=\int_0^{2 \pi} {d q \over 2 \pi}\; \sigma (q)
    \eu^{\ii q(j-k)}
 \label{Tg}
\ee
and in our case can be found from (\ref{F},\ref{G}) as
\be
   \sigma(q) = {1 \over 2}\left( 1 + \eu^{\ii\vartheta_q} \right)
   = {1 \over 2} + {\cos q - h + \ii \gamma \sin q
   \over 2 \sqrt{(\cos q-h)^2+\gamma^2\sin^2q}} .
   \label{genfunc}
\ee

Thus, the problem of calculation of the EFP
\be
   P(n) =  \left|\det ( {\bf S_n} ) \right|,
  \label{PnX}
\ee
is reduced (exactly) to the calculation of the determinant of
the $n \times n$ Toeplitz matrix ${\bf S_n}$ defined by the
generating function (\ref{Tg},\ref{genfunc}).
The representation (\ref{PnX}) is exact and valid for any $n$.
In our study we are interested in finding an asymptotic behavior of
(\ref{PnX}) at large $n \to \infty$. \footnote{The reader might notice that our
generating function (\ref{genfunc}) is almost the same as the one
analyzed by Barouch et al. in \cite{mccoy} ($\sigma_{[13]}(q)
= {\cos q - h + \ii \gamma\sin q \over \sqrt{ (\cos q - h)^2 + \gamma^2 \sin^2 q } } $).
The only difference is the shift by the unity in our expression.
This difference changes dramatically the analytical structure of the generating function, in particular, its winding
number around the origin, and requires a new analysis of the generated Toeplitz determinants.}

Most of these results are derived using known theorems on the
asymptotic behavior of Toeplitz determinants. We collect these
theorems in Appendix~\ref{ToeplitzApp}. In the following sections we apply
them to extract the corresponding asymptotes of $P(n)$ at $n \to
\infty$ in the different regions of the phase diagram.
Two major distinctions have to be made in this process.
For the critical isotropic ($\gamma = 0$) XY model, one applies
what is known as Widom's Theorem and one finds a Gaussian behavior
with a power law prefactor \cite{shiroishi}.
In the rest of the phase diagram, we apply different formulations
of what is known in general as the Fisher-Hartwig conjecture,
which always leads to an exponential asymptotic behavior.
As expected, we find a pure exponential decay for the EFP in the
non-critical regions.

For $h > 1$, the exponential decay is modulated by an additional 
oscillatory behavior.

At the critical magnetizations $h = \pm 1$, we discover an exponential
decay with a {\it power law pre-factor}.
Moreover, by extending the existing theorems on Toeplitz
determinants beyond their range of applicability, for $h = \pm 1$ we
obtain the first order corrections to the asymptotics as a faster
decaying power law with the same exponential factor.
For $h = 1$, the first order correction is also oscillating and this
means that the EFP presents an oscillatory behavior of the EFP for
$h \ge 1$.

The reader who is not interested in the mathematical details of our
derivations can find the results in Table \ref{table1} and skip the
following sections to go directly to Sec.~\ref{Crossover}, where we
analyze the crossover between the Gaussian behavior
at $\gamma = 0$ and the asymptotic exponential decay at finite
$\gamma$ using a bosonization approach.

\begin{table}
 \centering
   \noindent\begin{tabular}{|c|c|l|l|l|l|}
     \hline
     \multicolumn{6}{|c|}{\bfseries EFP for the Anisotropic XY model} \\
     \hline
      Region  &  $\gamma$, $h$  &  $P(n)$  &  Eq.  & Section & Theorem \\
     \hline
     \hline
      $\Sigma_-$  &  $h<-1$  &  $ E\, \eu^{-n \beta}$  &  \ref{pnc}  &  \ref{SigmaMSec}  &  Szeg\"o \\
     \hline
      $\Sigma_0$  &  $-1<h<1$  & $ E\, \eu^{-n \beta }$  &  \ref{pnc}  &  \ref{Sigma0Sec}  &  FH  \\
     \hline
      $\Sigma_+$  &  $h>1$  &
       $ {\it E} \left[ 1 + (-1)^n A \right] \: \eu^{-n \beta}$  &  \ref{oscbehavior}  &  \ref{SigmaPSec}  &  gFH \\
     \hline
      $\Gamma_E$  &  $\gamma^2+h^2=1$  & $ E\, \eu^{-n \beta}$  &  \ref{pnexact}  &  \ref{GammaESec}  &  Exact \\
     \hline
     \hline
      ${\bf \Omega_+}$  &  $h=1$  & 
       $ {\it E} \: n^{- {1 / 16} } \left[ 1 + (-1)^n A/\sqrt{n} \: \right] \: \eu^{- n \beta}$  &  \ref{pnomp}  &  \ref{ssOmegap}  &  gFH  \\
     \hline
      ${\bf \Omega_-}$  &  $h=-1$  & $ {\it E} \: n^{- {1 / 16} } \left[ 1 +  A/\sqrt{n} \: \right]  \: \eu^{- n \beta}$  &  \ref{pnomm}  &  \ref{ssOmegam}  &  gFH \\
     \hline
      ${\bf \Omega_0}$  &  $\gamma=0$,\, $|h|<1$  & $ E\, n^{-1/4}e^{-n^2 \alpha}$  &  \ref{pnsh}  &  \ref{gammazero}  & Widom  \\
     \hline
   \end{tabular}
   \caption{Asymptotic behavior of the EFP in different regimes.
    The exponential decay rate $\beta$ is given by Eq. (\ref{betagh})
    for all regimes. The regions in boldface are the critical ones.
    The coefficients $E,A$ are functions of $h$ and $\gamma$ 
    whose explicit expressions are provided in the text.
    Relevant theorems on Toeplitz determinants are collected in the
    \ref{ToeplitzApp}.}
   \label{table1}
\end{table}

%%%%%%%%%%%%%%%%%%%%%%%%%%%%%%%%%%%%%%%%%%%%%%%%%%%%%%%%%%%%%%%%%%
\section{Singularities of $\sigma(q)$ and exponential behavior of the
EFP}
\label{expsection}
%%%%%%%%%%%%%%%%%%%%%%%%%%%%%%%%%%%%%%%%%%%%%%%%%%%%%%%%%%%%%%%%%%%

To derive the asymptotic behavior of the EFP we rely on the
theorems on determinants of Toeplitz matrices.
These theorems depend greatly on the analytical structure of the
generating function (\ref{genfunc}), especially on its zeros and
singularities.

Setting $\gamma = 0$  in (\ref{genfunc}), we see that for the
Isotropic XY model the generating function has only a limited
support within its period $[0,2\pi]$.
This case is covered by what is known as Widom's Theorem and
will be considered in Section \ref{gammazero}.

\begin{figure}
   \dimen0=\textwidth
   \advance\dimen0 by -\columnsep
   \divide\dimen0 by 3
   \noindent\begin{minipage}[t]{\dimen0}
   \begin{flushleft}
   (a): ${\bf \Sigma_-}$
   \end{flushleft}
   \includegraphics[width=\columnwidth]{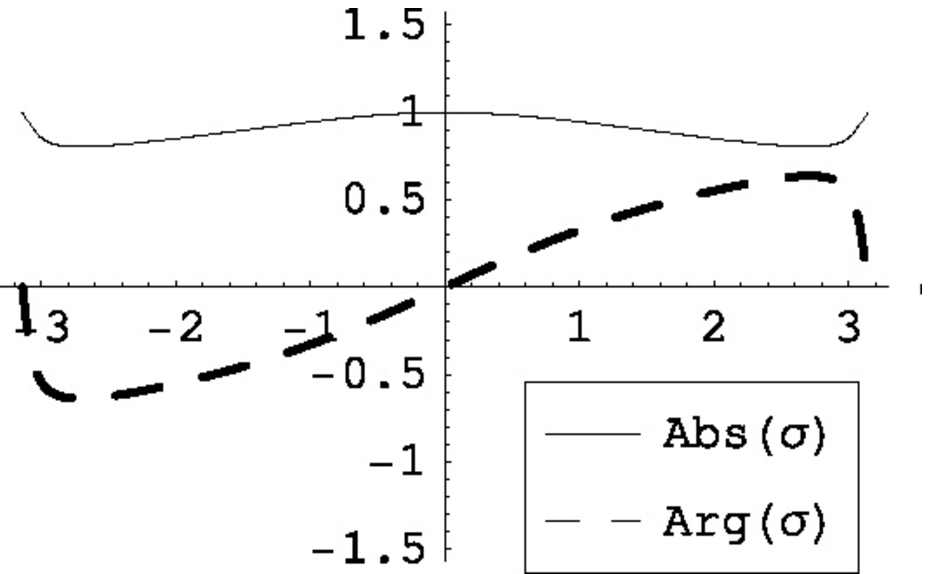}
   \end{minipage}
   \hfill
   \begin{minipage}[t]{\dimen0}
   \begin{flushleft}
   (b): ${\bf \Omega_-}$
   \end{flushleft}
   \includegraphics[width=\columnwidth]{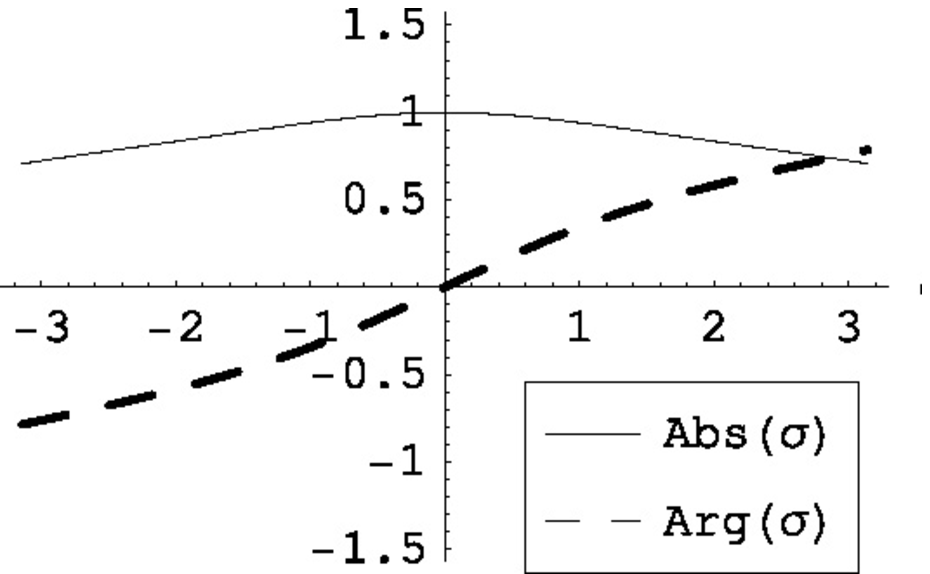}
   \end{minipage}
   \hfill
   \begin{minipage}[t]{\dimen0}
   \begin{flushleft}
   (c): ${\bf \Sigma_0}$
   \end{flushleft}
   \includegraphics[width=\columnwidth]{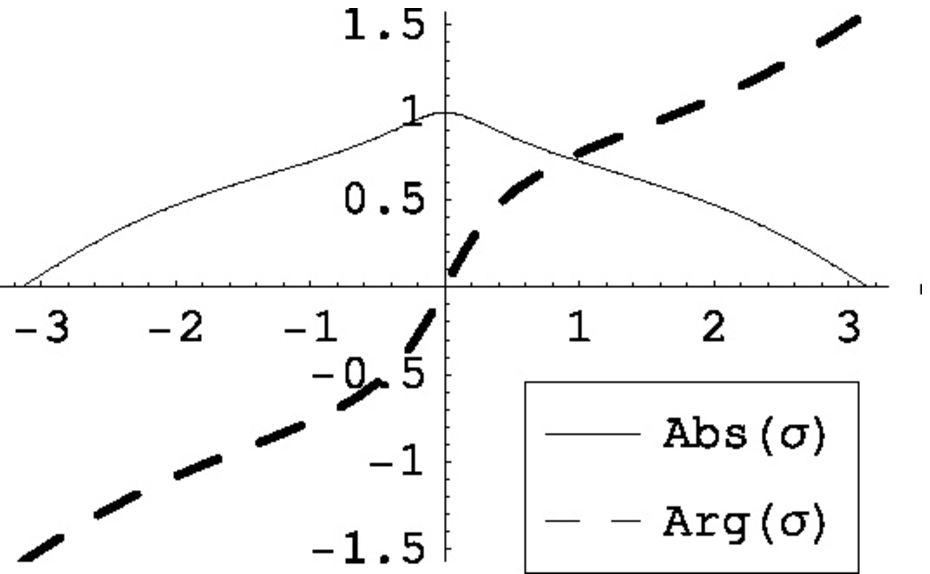}
   \end{minipage}
   \noindent\begin{minipage}[t]{\dimen0}
   \begin{flushleft}
   (d): ${\bf \Omega_+}$
   \end{flushleft}
   \includegraphics[width=\columnwidth]{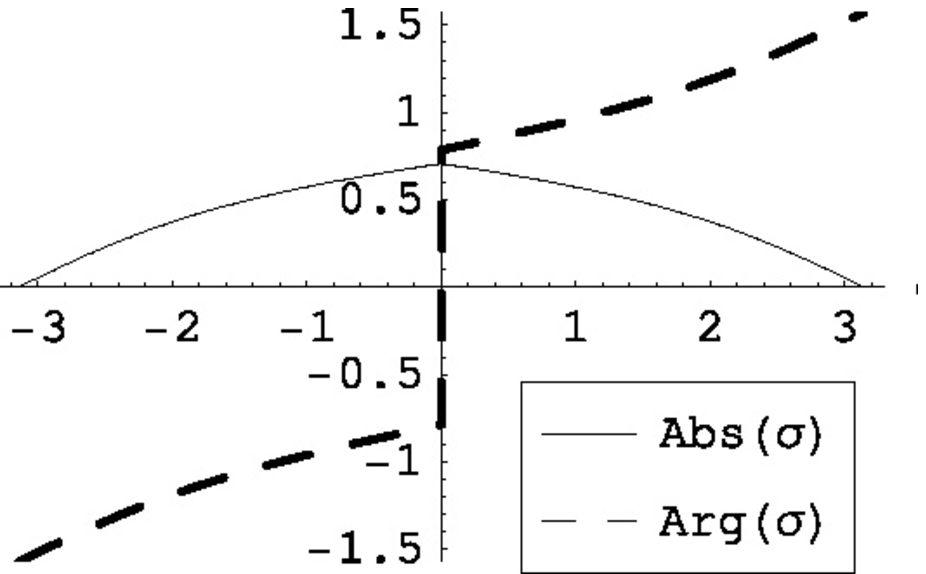}
   \end{minipage}
   \hfill
   \begin{minipage}[t]{\dimen0}
   \begin{flushleft}
   (e): ${\bf \Sigma_+}$
   \end{flushleft}
   \includegraphics[width=\columnwidth]{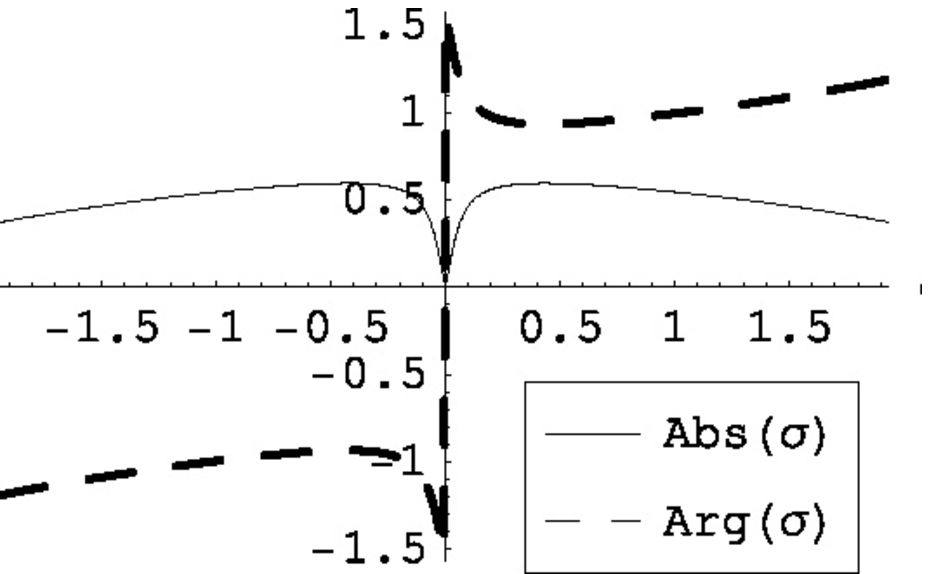}
   \end{minipage}
   \hfill
   \begin{minipage}[t]{\dimen0}
\caption{Plot of the absolute value and argument of the generating
function (\ref{genfunc}) for $\gamma = 1.5$ at different values of
$h$. From (a) to (e) $h = -1.1$, $-1$, $0.5$, $1$, $1.1$, respectively.}
   \label{GenFuncPlot}
   \end{minipage}
\end{figure}

In the remaining parts of the phase-diagram the generating function has
only pointwise singularities (zeros) as it is shown in
Fig.~\ref{GenFuncPlot}.
These cases are treated under a general (not yet completely proven)
conjecture known as the Fisher-Hartwig conjecture (FH), which
prescribes the leading asymptotic behavior of the Toeplitz determinant
to be exponential in $n$:
\be
   P(n) {\stackrel{n \rightarrow \infty}{\sim}} \eu^{-\beta n}.
\ee

While the pre-exponential factors depend upon the particulars of the
singularities of the generating function, the exponential decay rate is
given in the whole phase diagram ($\gamma \neq 0$) according to FH as
\bea
   \beta(h, \gamma) & = &
   - \int_0^{2 \pi} {\de q \over 2 \pi}\; \log  \left| \sigma(q) \right|
   \nonumber \\ & = &
   - \int_0^\pi {\de q \over 2 \pi}\;
   \log \left[ {1 \over 2} \left( 1 + {\cos q -h \over \sqrt{
   \left( \cos q - h \right)^2 + \gamma^2  \sin^2 q} } \right) \right] .
   \label{betagh}
\eea
The integral in (\ref{betagh}) is convergent for all $h$ and all
$\gamma \neq 0$ and $\beta(h,\gamma)$ is a continuous function of its
parameters.

\begin{figure}
  \includegraphics[width=\columnwidth]{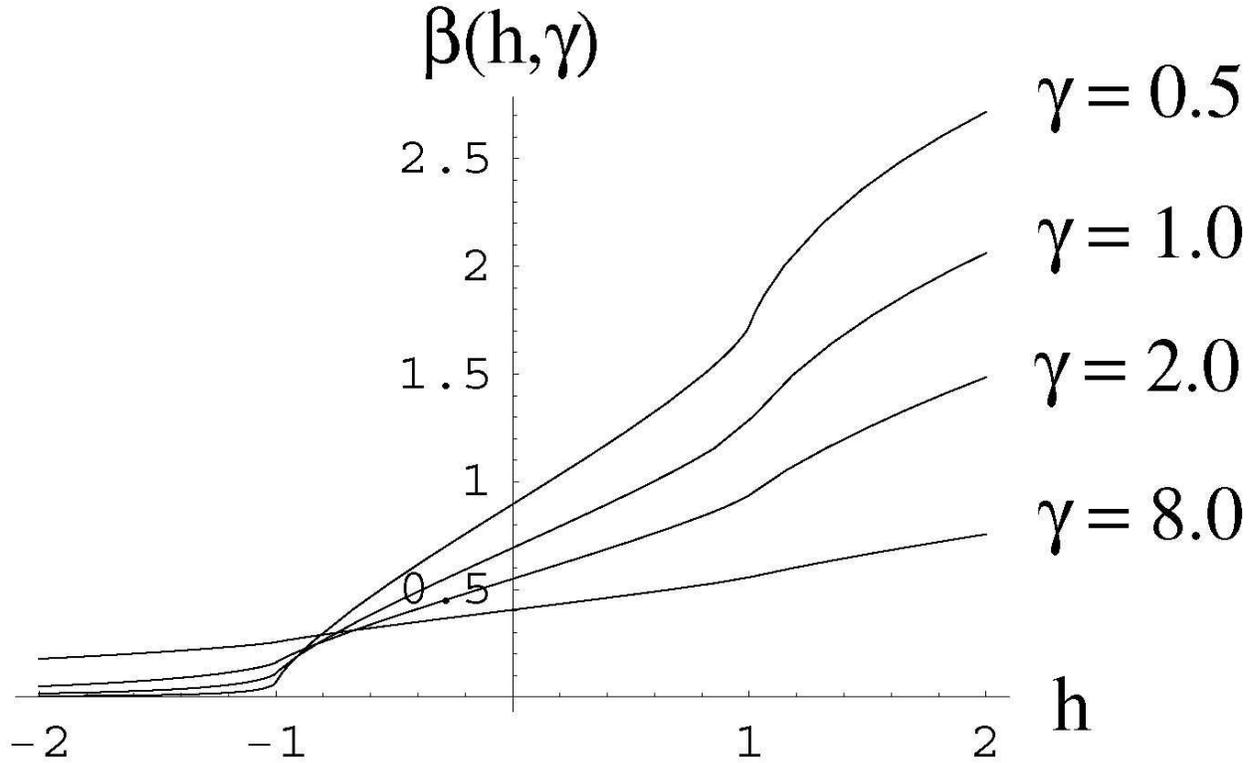}
\caption{Plot of the decay rate $\beta$ as a function of the
parameters $\gamma$ and $h$. The function diverges for $\gamma = 0$
and is continuous for $h = \pm 1$ (although has weak singularities at
$h=\pm 1$).}
   \label{betagraph}
\end{figure}

In Fig.~\ref{betagraph}, $\beta(h,\gamma)$ is plotted as a function
of $h$ at several values of $\gamma$.
One can see that $\beta(h,\gamma)$ is continuous but has weak
(logarithmic) singularities at $h= \pm 1$.
This is one of the effects of the criticality of the model on
the asymptotic behavior of EFP.

These weak singularities are also a manifestation of the rich analytical
structure underlying $\beta(h,\gamma)$ and the generating function
(\ref{genfunc}).
To understand these structures, we interpret the periodic generating
function (\ref{genfunc}) as the restriction to the unit circle
($z= \eu^{\ii \theta}$) of the complex function
\be
   \sigma (z) \equiv {1 \over 2} \left( 1 +
   { p_1 (z) \over \sqrt{p_1 (z) \cdot p_2 (z)} } \right),
   \label{sigmaz}
\ee
where
\bea
   p_1 (z) = {1 + \gamma \over 2 z} (z - z_1) (z - z_2), \\
   p_2 (z) = {1 + \gamma \over 2 z} (z_1 z - 1) (z_2 z - 1)
\eea
with
\bea
   z_1 = {h - \sqrt{h^2 + \gamma^2 - 1} \over 1 + \gamma},
   \label{z1} \\
   z_2 = {h + \sqrt{h^2 + \gamma^2 - 1} \over 1 + \gamma}.
   \label{z2}
\eea

\begin{figure}
   \dimen0=\textwidth
   \advance\dimen0 by -\columnsep
   \divide\dimen0 by 3
   \noindent\begin{minipage}[t]{\dimen0}
   (a)  
   \includegraphics[width=\columnwidth]{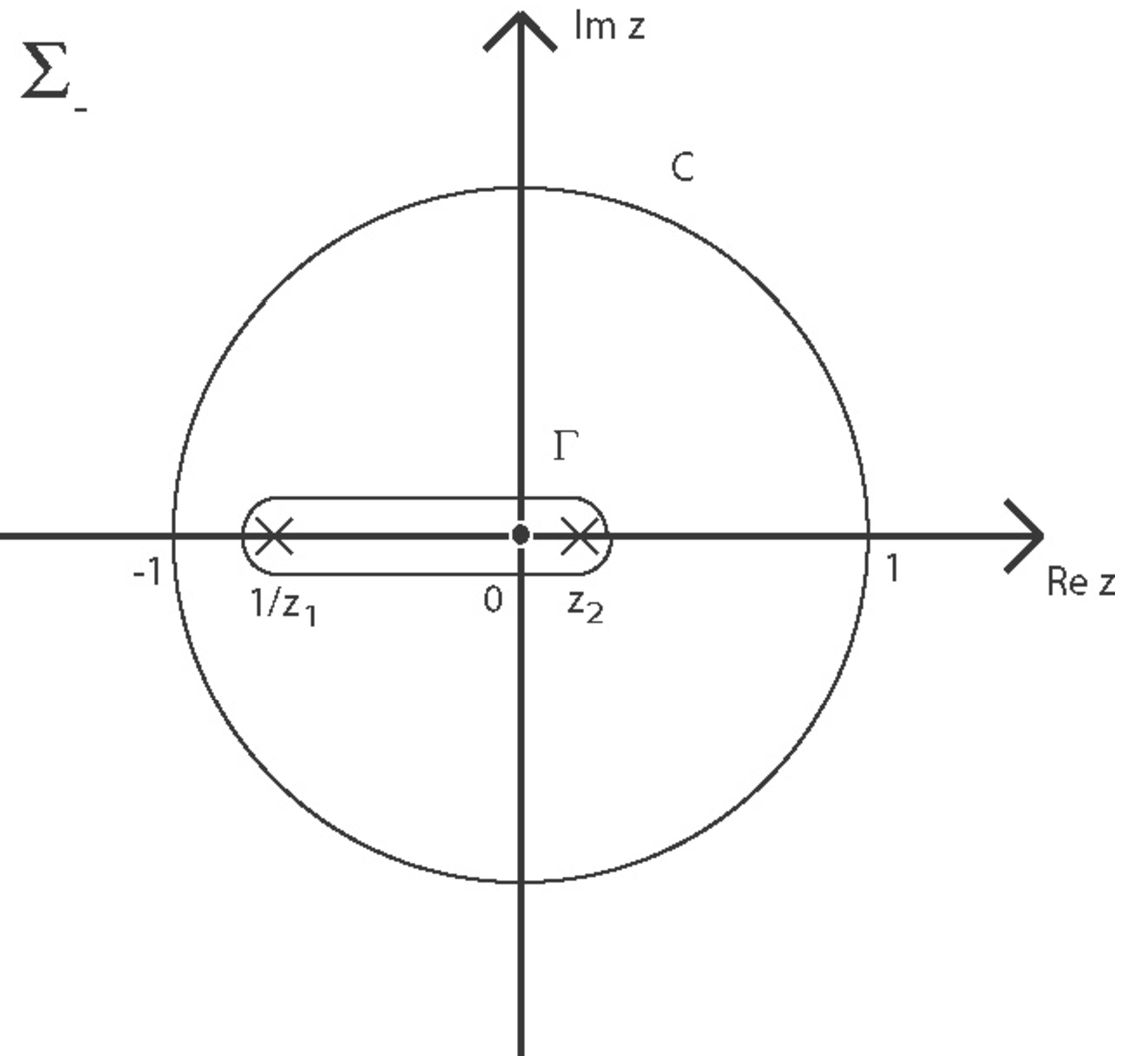}
   \end{minipage}
   \hfill
   \begin{minipage}[t]{\dimen0}
   (b)
   \includegraphics[width=\columnwidth]{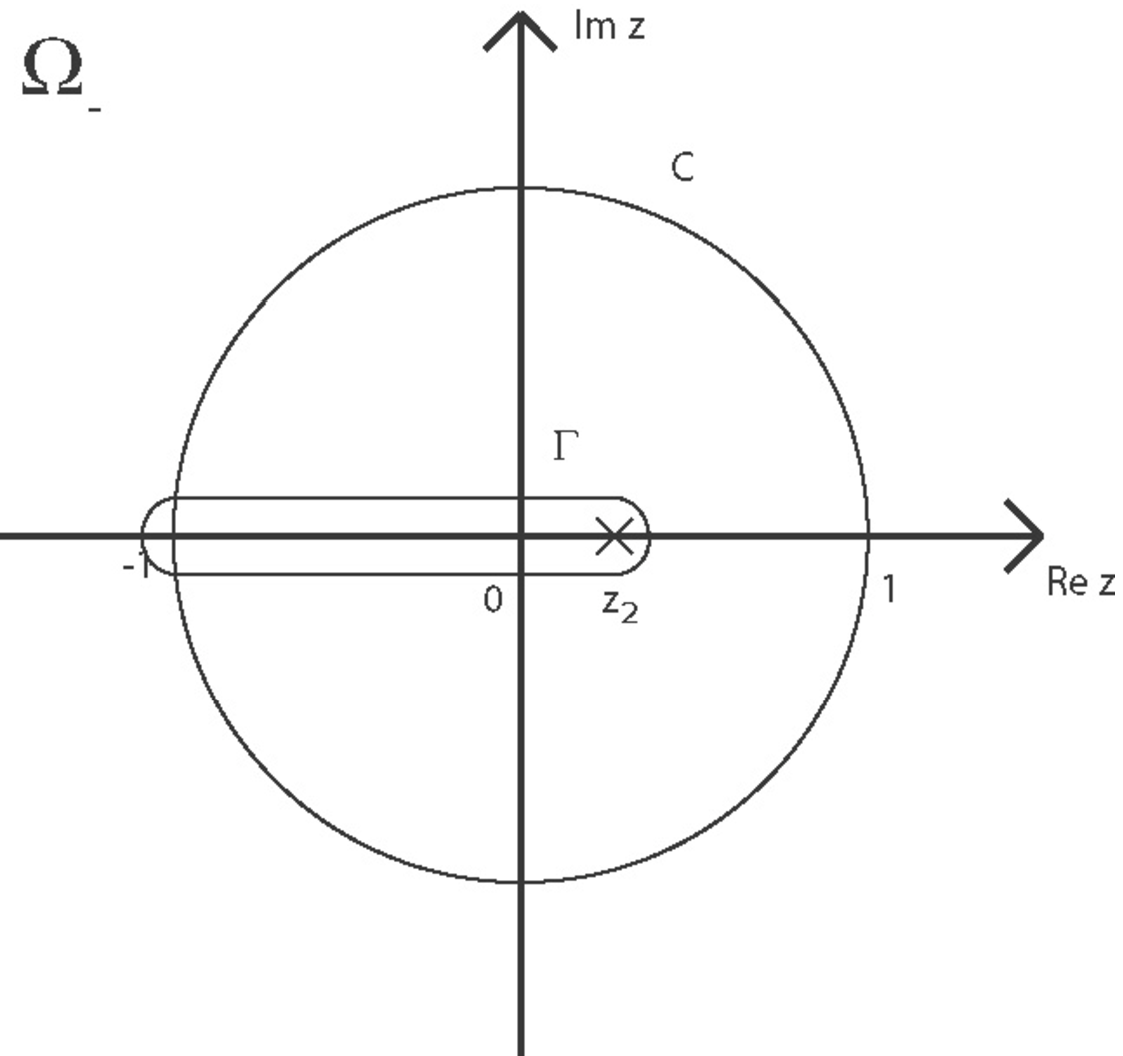}
   \end{minipage}
   \hfill
   \begin{minipage}[t]{\dimen0}
   (c)
   \includegraphics[width=\columnwidth]{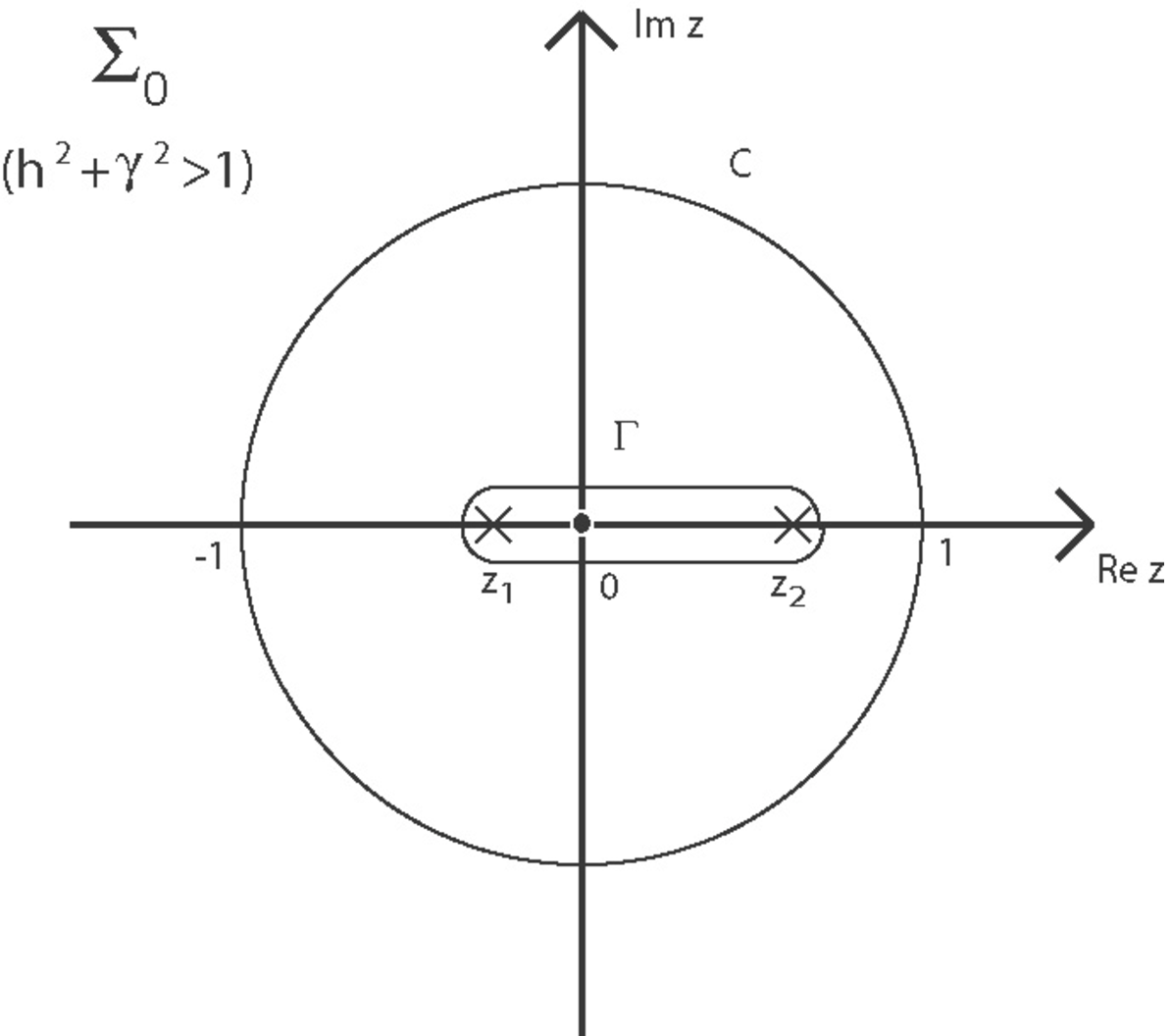}
   \end{minipage}
   \noindent\begin{minipage}[t]{\dimen0}
   (d)
   \includegraphics[width=\columnwidth]{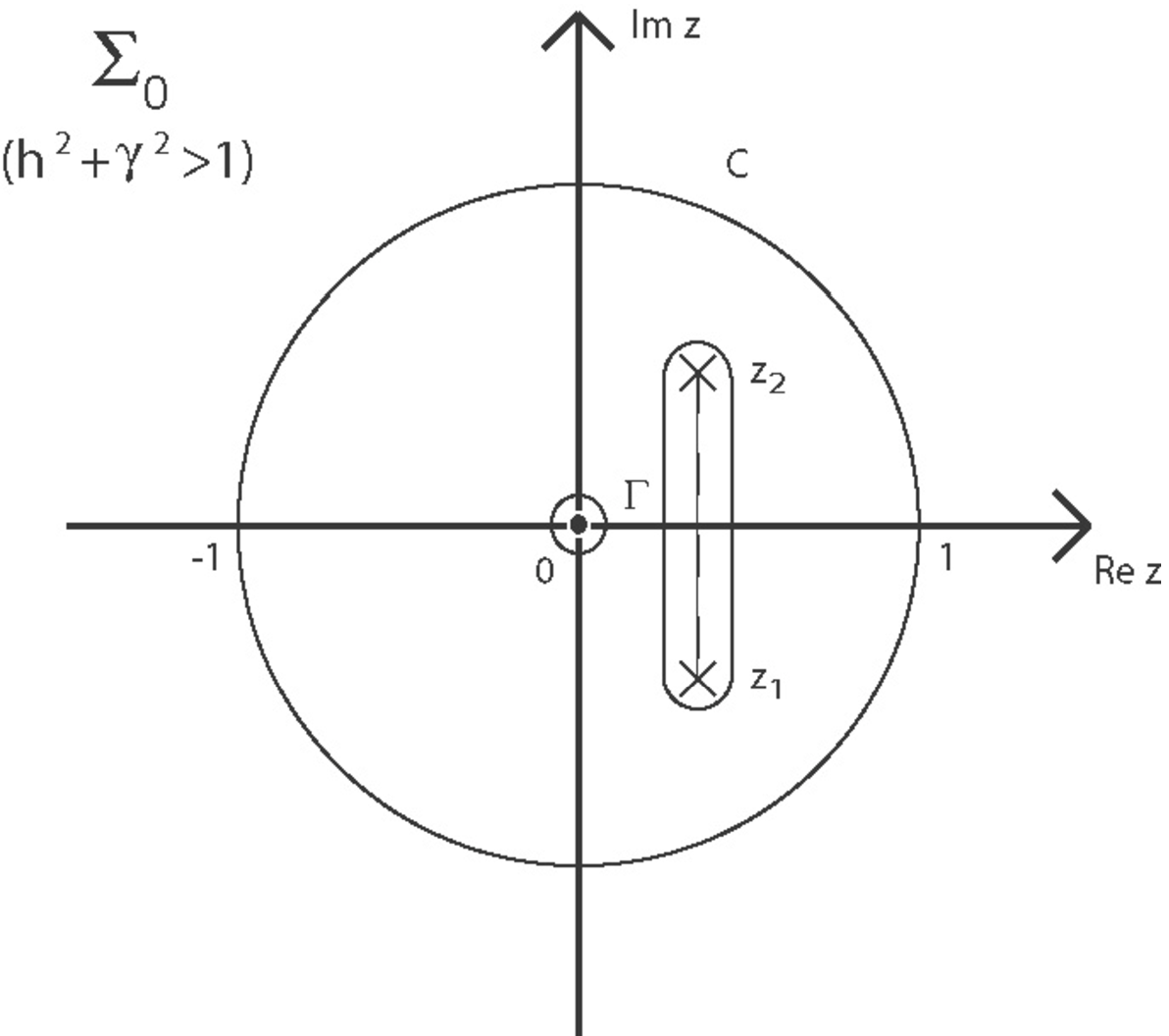}
   \end{minipage}
   \hfill
   \begin{minipage}[t]{\dimen0}
   (e)
   \includegraphics[width=\columnwidth]{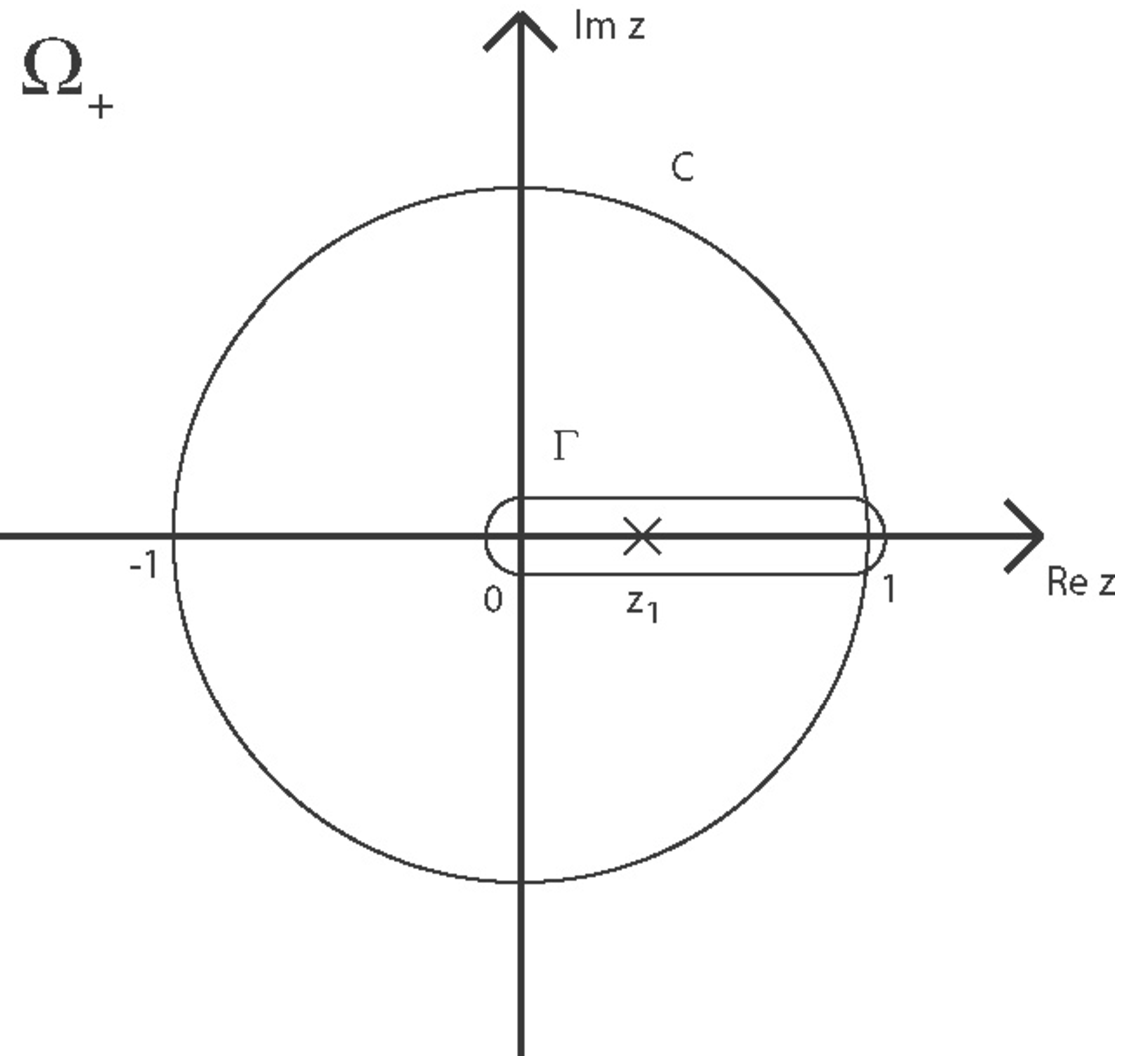}
   \end{minipage}
   \hfill
   \begin{minipage}[t]{\dimen0}
   (f)
   \includegraphics[width=\columnwidth]{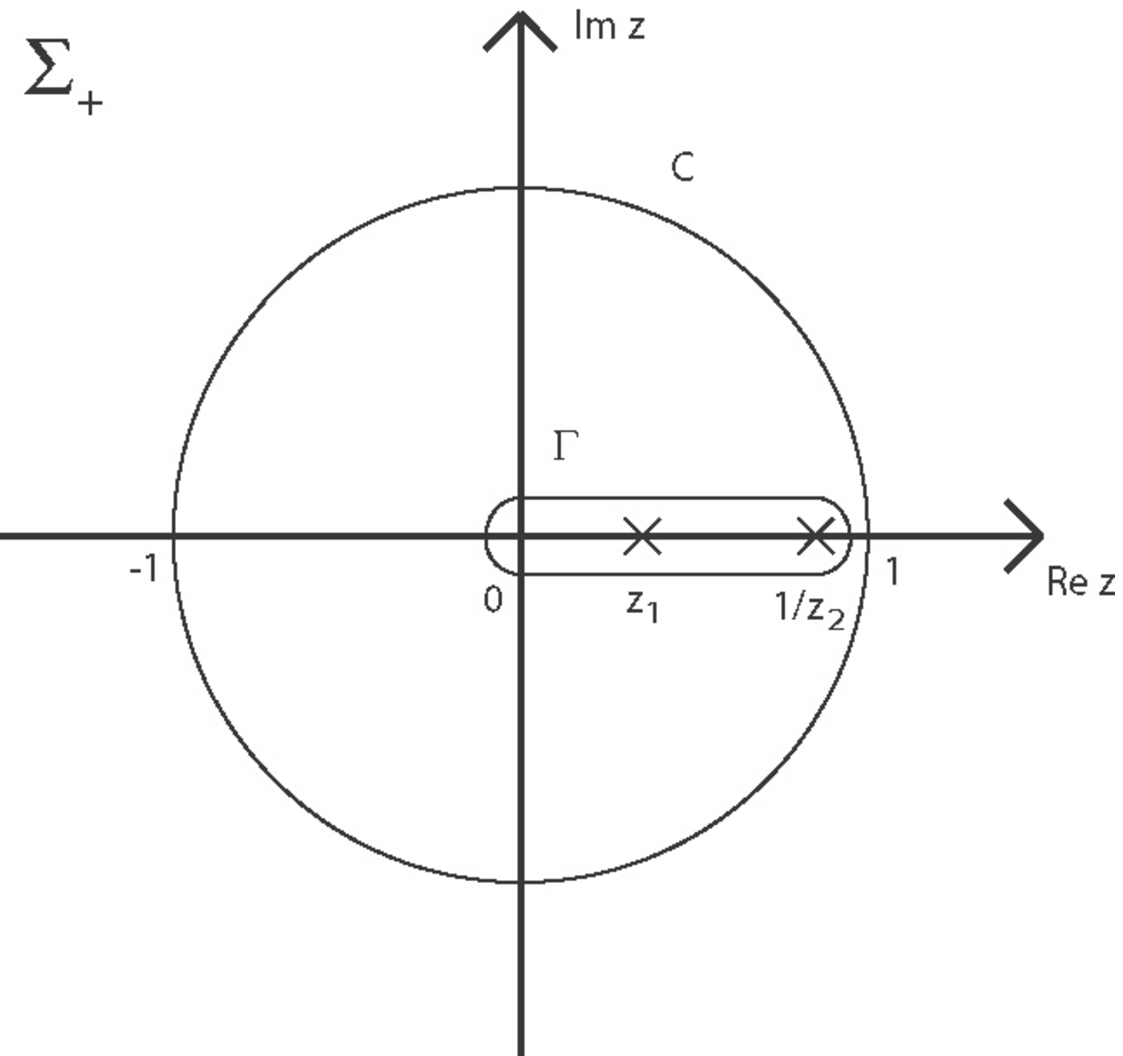}
   \end{minipage}
   \begin{minipage}[t]{\textwidth}
\caption{The integral in (\ref{betagh}) is performed over the unit circle
$C$. The analytical structure of the integrand allows for a deformation
of the contour of integration into $\Gamma$, which encloses a
logarithmic branching line, different in the various regions of the
phase-diagram (in (d), $\Gamma$ encloses also a simple pole at
the origin). The roots $z_1$ and $z_2$ were defined in (\ref{z1})
and (\ref{z2}).}
   \label{SigmaStruct}
   \end{minipage} 
\end{figure}

The integral in (\ref{betagh}) can be regarded as a contour integral 
over the unit circle of the function (\ref{sigmaz}). We can deform the
contour of integration taking into account the complex structure of the
integrand in the various regions (see Fig.~\ref{SigmaStruct}) and
express (\ref{betagh}) as a simpler integral on the real axis (after
partial integration and some algebra).

%%%%%%%%%%%%%%%%%%%%%%%%%%%%%%%%%%%%%%%%%%%%%%%%%%%%%%%%%
\subsection{The non-critical regions ($\Sigma_\pm$ and $\Sigma_0$)}

%%%%%%%%%%%%%%%%%%%%%%%%%%%%%%%%%%%
\subsubsection{$\Sigma_-$ ($h < -1$)}

For $h<-1$, the analytical structure of the integrand of (\ref{betagh})
is shown in Fig.~\ref{SigmaStruct}a.
We re-write the decay rate (\ref{betagh}) in this region as
\be
   \beta (h,\gamma)  = {1 \over 2}
   \ln \left[ {\sqrt{h^2 + \gamma^2 -1} - h \over \gamma +1 } \right]
   - \Lambda(h,\gamma) - \Delta(h,\gamma),
   \label{betaminus}
\ee
where
\bea
   \Lambda(h,\gamma) & \equiv & \ln \left| {1 \over 2} \left( 1 - {h \over |h|}
   \sqrt{ {1 - \gamma \over 1 + \gamma} } \right) \right|,
   \label{Lambda} \\
   \Delta(h,\gamma) & \equiv & \int^1_{|K|} {\de x \over 2 \pi}
   {1 \over \sqrt{(1-x^2) (x^2 - K^2)} }
   \left( x + {K \over x} \right)
   \ln \left| {x- a \over x + a } \right|
   \label{Delta}
\eea
with
\bea
   K & \equiv & { \sqrt{h^2 + \gamma^2 -1} - \gamma
   \over \sqrt{h^2 + \gamma^2 -1} + \gamma },
   \label{kappa}
   \\
   a & \equiv &
   {\sqrt{h^2 + \gamma^2 -1} - \gamma \over h - 1 } .
\eea
This decomposition of $\beta (h,\gamma)$ is especially useful
in analyzing the transitions between non-critical and critical
regimes.  In fact, we will see that the functions
$\Lambda(h,\gamma)$ and $\Delta(h,\gamma)$ defined above are
universal across the phase diagram (hence the need for the seemingly
redundant absolute values in our definitions).
%The absolute value in the logarithm of the integrand is relevant for $\gamma > 1$,
%since its argument changes sign and vanishes within the interval of integration
%($a>K$ for $\gamma >1$).

%%%%%%%%%%%%%%%%%%%%%%%%%%%%%%%%%%%
\subsubsection{$\Sigma_0$ ($|h| < 1$)}

As before, we can express the contour integral defining
$\beta(h,\gamma)$ as a standard integral on the real axis.
For $|h| < 1$ and $h^2 + \gamma^2 > 1$, the structure of the integrand
is depicted in Fig.~\ref{SigmaStruct}c and the decay rate is simply
\be
   \beta(h,\gamma) =  - \Lambda(h,\gamma) - \Delta(h,\gamma),
   \label{beta0}
\ee
where $\Lambda(h,\gamma)$ and $\Delta(h,\gamma)$ have already been
defined in (\ref{Lambda},\ref{Delta}).

For $h^2 + \gamma^2 < 1$, the structure is quite different (see
Fig.~\ref{SigmaStruct}d). In this region the expression for
$\beta(h,\gamma)$ in terms of a real axis integral is complicated and
will therefore be omitted it in this paper.

\subsubsection{$\Sigma_+$ ($h > 1$)}

A calculation similar to the previous ones (see Fig.~\ref{SigmaStruct}f)
gives the expression for the decay factor for $h > 1$:
\be
   \beta (h,\gamma)  = {1 \over 2}
   \ln \left[ {\sqrt{h^2 + \gamma^2 -1} + h \over \gamma + 1 } \right]
   - \Lambda(h,\gamma) - \Delta(h,\gamma),
   \label{betaplus}
\ee
where $\Lambda(h,\gamma)$ and $\Delta(h,\gamma)$ were introduced in
(\ref{Lambda}) and (\ref{Delta}).

One important difference exists in this region: as will be
discussed in length later in Section \ref{SigmaPSec}, in $\Sigma_+$
there are two equivalent representations of the generating function.
This ambiguity reflects on the value of $\beta$, in that the choice of
the representation for the generating function determines the branch cuts
in Fig.~\ref{SigmaStruct}.
We will see that we have to use both values of $\beta$, which differ
only by an imaginary constant:
\be
   \beta' = \beta + \ii \pi
   \label{betapluspi}
\ee
and this will add an oscillatory behavior to the EFP.

%%%%%%%%%%%%%%%%%%%%%%%%%%%%%%%%%%%%%%%%%%%%%%%%%%%%%%%%%
\subsection{The critical lines ($\Omega_\pm$)}

We can calculate the decay factor $\beta$ at $h=1$ ($\Omega_+$) from a
limiting procedure on (\ref{beta0}) or (\ref{betaplus}).
At $h = 1$, only $\Delta(h,\gamma)$ is nonvanishing, thus guaranteeing
the continuity of $\beta$ across the critical line.
From an appropriate limit of (\ref{Delta}), we calculate the decay
rate for $h = 1$ as
\be
   \beta(1,\gamma) = - \int_0^1 {\de x \over 2\pi}
   {1 \over \sqrt{1 - x^2} }
   \ln \left| {1 -\gamma x \over 1 + \gamma x} \right|
   - \ln \left| {1 \over 2}
   \left( 1 - \sqrt{ 1 - \gamma \over 1 + \gamma} \right) \right| .
   \label{betaPlus}
\ee
%For $\gamma <1$, we can expand the logarithm in series and perform the
%integral:
%\be
%   \beta(1,\gamma <1) = {1 \over \sqrt{\pi} }
%   \sum_{n=0}^\infty { n! \over \Gamma(n + 1/2) }
%   { \gamma^{2 n + 1} \over (2 n + 1)^2 }
%    - \ln \left[ {1 \over 2}
%   \left( 1 - \sqrt{ 1 - \gamma \over 1 + \gamma} \right) \right].
%\ee
As discussed before in reference to (\ref{betaplus}), the definition of
$\beta(1,\gamma)$ is not unique and, as in (\ref{betapluspi}),
will generate again an oscillatory behavior for the EFP (see later in
Sec.~\ref{ssOmegap}).

The value of $\beta$ at $h = -1$ can also be obtained  from a limiting
procedure on (\ref{Delta})
\be
   \beta(-1,\gamma) = \int_0^1 {\de x \over 2\pi}
   {1 \over \sqrt{1 - x^2} } \ln \left| {1 -\gamma x \over 1 + \gamma x} \right|
   - \ln \left| {1 \over 2} \left( 1 + \sqrt{ 1 - \gamma \over 1 + \gamma} \right)
   \right| .
   \label{betaMinus}
\ee
%Again, for $\gamma <1$, we can expand the logarithm in series to calculate the
%integral:
%\be
%   \beta(-1,\gamma <1) = - {1 \over \sqrt{\pi} }
%   \sum_{n=0}^\infty { n! \over \Gamma(n + 1/2) }
%   { \gamma^{2 n + 1} \over (2 n + 1)^2 }
%    - \ln \left[ {1 \over 2}
%   \left( 1 + \sqrt{ 1 - \gamma \over 1 + \gamma} \right) \right].
%\ee

As can be seen in Fig.~\ref{betagraph}, the decay factor $\beta$ is
continuous across the critical lines, but has a discontinuity in its
derivative.
As $\beta$ approaches the critical lines, it actually shows a
non-analytical behavior leading to a logarithmic singularity:
\be
   \beta(h=\pm 1 + \epsilon,\gamma) = \beta(\pm 1,\gamma)
   + {\gamma \over \pi} \; \epsilon \ln | \epsilon | .
\ee
The derivative $d\beta/dh$ diverges logarithmically as
$h \to \pm 1$.

Moreover, one can easily notice from the difference between expression
(\ref{beta0}) and (\ref{betaplus}) that even the finite part of
the derivative of $\beta(h,\gamma)$ by $h$ is different if one
approaches the critical line $h = 1$ from above or below, due to the
appearance of the additional term in (\ref{betaplus}).
The same holds across the critical line $h = -1$, due to the presence
of the first term in (\ref{betaminus}), which doesn't appear in
(\ref{beta0}).

%%%%%%%%%%%%%%%%%%%%%%%%%%%%%%%%%%%%%%%%%%%%%%%%%%%%%%%%%%%%%%%%%%%
\section{The pre-exponential factors}
\label{pre-ex}
%%%%%%%%%%%%%%%%%%%%%%%%%%%%%%%%%%%%%%%%%%%%%%%%%%%%%%%%%%%%%%%%%%%

For $\gamma \ne 0$, the leading behavior of the EFP is always
exponential. However, the singularities of  the generating function
are different in different regions of the phase diagram and we must
therefore use different forms of the Fisher-Hartwig
conjecture in order to derive the pre-exponential factors and determine
the asymptotic behavior of $P(n)$.
We will now show how to obtain the results for each of the regions.

%%%%%%%%%%%%%%%%%%%%%%%%%%%%%%%%%%%%%%%%
\subsection{The non-critical regions ($\Sigma_\pm$ and $\Sigma_0$)}
\label{non-critical}
%%%%%%%%%%%%%%%%%%%%%%%%%%%%%%%%%%%%%%%%

%%%%%%%%%%%%%%%%%%
\subsubsection{$\Sigma_-$ ($h < -1$)}
\label{SigmaMSec}

In this region ($\gamma \neq 0$, $h<-1$) the generating function
(\ref{genfunc}) is nonzero for all $q$ (see Fig.~\ref{GenFuncPlot}a):
this is the simplest case and can be treated using the (rigorously
proven) {\it Strong Szeg\"o Limit Theorem}, see (\ref{szego}).
It gives
\be
  P(n) =  \left| \det ({\bf S_n}) \right|
  {\stackrel{n \rightarrow \infty}{\sim}}
  {\it E_-} (h,\gamma) \eu^{-\beta(h,\gamma) n}
  \label{pnc}
\ee
with $\beta(h,\gamma)$ given by (\ref{betagh}) or (\ref{betaminus}) and
\be
   {\it E_-} (h, \gamma) =
   \exp\left(\sum_{k=1}^\infty  k \hat{\sigma}_k \hat{\sigma}_{-k}\right),
   \label{Esgh}
\ee
where $\hat{\sigma}_k$ is defined in (\ref{sigmak}) as the $k$-th
Fourier component of the logarithm of $\sigma$:
\bea
  \hat{\sigma}_k & \equiv & \int_0^{2 \pi} {\de q \over 2 \pi}\;
  \left[ \log \sigma(q) \right] \eu^{- \ii k q}
  \nonumber \\
  & = &
  \int_0^{2 \pi} {\de q \over 2\pi}\; \eu^{-\ii k q}
   \log \left( 1 + {\cos q - h + \ii \gamma \sin q \over
   \sqrt{ \left( \cos q - h \right)^2 + \gamma^2  \sin^2 q}} \right).
  \label{sigmahat}
\eea
The sum in (\ref{Esgh}) is convergent only for $\gamma \ne 0$ and for
$h < - 1$.
For $h \ge -1$, the generating function (\ref{genfunc}) develops
singularities which produce $1/k$ contributions to
(\ref{sigmahat}) that make the sum in (\ref{Esgh}) divergent.
Therefore, in the rest of the phase diagram these singularities
have to be treated to absorb the harmonic series contributions.
Consequently, each region of the phase diagram will involve a different
definition for the pre-exponential factor and the "regularization"
procedure will sometimes generate an additional power-law contribution.
The result is given by  the Fisher-Hartwig conjecture that we must use
in the remainder of the phase diagram.

%%%%%%%%%%%%%%%%%%%%%%%%%%%%%%%%%%%%%%%%%%%
\subsubsection{$\Sigma_0$ ($|h| < 1$)}
\label{Sigma0Sec}

As can be noticed from Fig.~\ref{GenFuncPlot}c, in $\Sigma_0$
($\gamma\neq 0$, $-1<h<1$) the generating function $\sigma(q)$ vanishes
and  its phase has a discontinuity of $\pi$ at $q = \pi$.
The asymptotic behavior of Toeplitz determinants with this type of
singularities in the generating function is given by FH, which is
actually proven for cases in which only one singularity is present.

We decompose the generating function as in (\ref{fishdec})
\be
   \sigma(q) = \tau(q) \eu^{{\ii \over 2} [(q - \pi) \mod 2 \pi - \pi]}
   \left( 2 - 2 \cos (q - \pi) \right)^{1/2}
   \label{param0}
\ee
and using (\ref{singexp}) we obtain
\be
   P(n) = \left| \det ({\bf S_n}) \right|
   {\stackrel{n \rightarrow \infty}{\sim}}
   {\it E_0 }(h,\gamma) \: \eu^{- \beta(h,\gamma) n}.
   \label{expbehavior}
\ee
The behavior is exponential as before with the  decay rate
$\beta(h,\gamma)$ from (\ref{betagh},\ref{beta0}), but the
pre-exponential factor is different. According to (\ref{fisherhartwig})
it is given by
\be
   {\it E_0 }(h,\gamma) \equiv { E[\tau] \over \tau_- (\pi) },
\ee
where, as in (\ref{szegoexp}) and (\ref{sigmak})
\bea
   E [\tau] & = &
   \exp\left(\sum_{k=1}^\infty  k \hat{\tau}_k \hat{\tau}_{-k}\right)
   \label{Etgh}
\eea
and
\be
   \hat{\tau}_k = \hat{\sigma}_k -  {(-1)^k \over k} \; \theta (k).
\ee
Here $\theta (k)$ is the usual Heaviside step function.
As we mentioned in the previous section, $\hat{\sigma}_k$ (\ref{sigmahat})
has $1/k$ contributions from singularities of $\sigma(q)$ and 
the effect of the parametrization (\ref{param0}) is to cure
(remove) these harmonic series divergences of the prefactor of the EFP
in this regime.

%%%%%%%%%%%%%%%%%%%%%%%%%%%%%%%%%%%%%%%%%%%%%%
\subsubsection{$\Sigma_+$ ($h > 1$)}
\label{SigmaPSec}

In $\Sigma_+$ ($\gamma\neq 0$, $h>1$), $\sigma(q)$ vanishes at $q = 0$
and $q = \pi$ and its phase presents two $\pi$ jumps at those points
(Fig.~\ref{GenFuncPlot}e).

In this case the application of FH leads to some ambiguity, because
there exist two representations of the kind (\ref{fishdec}) and one
obtains two values for $\beta(h,\gamma)$ using the two representations
of the generating function: $\beta_1=\beta$ and
$\beta_2 = \beta +\ii \pi$, with $\beta$ from (\ref{betagh}) or 
(\ref{betaplus}).
This ambiguity is resolved by the (yet unproven) generalized
Fisher-Hartwig conjecture (gFH), which gives EFP as a sum of two terms so that
both values of $\beta$'s are used (see the appendix \ref{gfhsec} or \cite{basor}).

The two leading inequivalent parametrizations (\ref{fishgendec}) are:
\bea
   \sigma(q) & = & \tau^1(q) \eu^{{\ii \over 2}
   [(q - \pi)  \mod 2 \pi - \pi]}
   \left( 2 - 2 \cos (q - \pi) \right)^{1/2} \nonumber \\
   && \qquad \times \eu^{- {\ii \over 2} [q \mod 2 \pi - \pi]}
   \left( 2 - 2 \cos q \right)^{1/2} \\
   & = & \tau^2(q) \eu^{- {\ii \over 2} [(q - \pi) \mod 2 \pi - \pi]}
   \left( 2 - 2 \cos (q - \pi) \right)^{1/2} \nonumber \\
   && \qquad \times \eu^{{\ii \over 2} [q \mod 2 \pi - \pi]}
   \left( 2 - 2 \cos q \right)^{1/2}.
\eea
Application of (\ref{singgenexp}) gives the asymptotic behavior of the
determinants as
\be
   \left| \det ({\bf S_n}) \right|
   {\stackrel{n \rightarrow \infty}{\sim}}
   \left[ {\it E^1_+ }(h,\gamma) + (-1)^n {\it E^2_+}(h,\gamma) \right]
   \: \eu^{- \beta(h,\gamma) n}
\ee
with
\bea
   {\it E^1_+ }(h,\gamma) \equiv
   { E [\tau] \over \tau_+ (0) \tau_- (\pi) }, \\
   {\it E^2_+ }(h,\gamma) \equiv
   { E [\tau] \over \tau_+(\pi) \tau_- (0) }
\eea
and $\beta(h, \gamma)$, $E [\tau]$ defined in
(\ref{betagh},\ref{Etgh}) with
\be
   \hat{\tau}_k = \hat{\sigma}_k
   - {(-1)^k \over k} \; \theta (k) - {1 \over k} \; \theta (-k).
\ee
Once again, as in the previous section, the effect of the parametrization
is to remove the $1/k$ contributions to $\hat{\sigma}_k$ (\ref{sigmahat}) 
due to the singularities of the generating function.

We  conclude that the non-critical theory presents an
exponential asymptotic behavior of the EFP. In the region
$\Sigma_+$, however, the EFP in addition has even-odd oscillations
\be
   P(n) {\stackrel{n \rightarrow \infty}{\sim}} {\it E^1_+}(h,\gamma)
   \left[ 1 + A_+ (h,\gamma) \cos (\pi n) \right] \: \eu^{- \beta(h, \gamma) n},
   \label{oscbehavior}
\ee
where the exponential decay factor is given by (\ref{betaplus}).

The amplitude of the oscillations is
\bea
   A_+ (h,\gamma) & \equiv &
   { \tau_+ (0) \tau_- (\pi) \over \tau_- (0) \tau_+ (\pi)} \nonumber \\
   & = & { \tau (0) \over \tau (\pi)} \left( {\tau_- (\pi)
   \over \tau_- (0)} \right)^2 \nonumber \\
   & = & { h + 1 \over h - 1 } \exp \left( 4 \lim_{\epsilon \rightarrow 0}
   \oint {\de z \over 2 \pi} {\log \tau (z) \over z^2 - (1 + \epsilon)^2} \right),
   \label{AplusIn}
\eea
where we used (\ref{wienerhopf}), the definition of $\tau$ and
(\ref{wienint}). 
We can deform the contour of integration as in Fig.~\ref{SigmaStruct}f
and calculate the integral in (\ref{AplusIn}) to obtain 
\be
   A_+ (h,\gamma) = \sqrt{K(h,\gamma)} =
   {\sqrt{h^2 - 1} \over \sqrt{h^2 + \gamma^2 - 1} + \gamma} \; ,
   \label{Aplus}
\ee
where $K(h,\gamma)$ was defined in (\ref{kappa}).

\begin{figure}
  \includegraphics[width=\columnwidth]{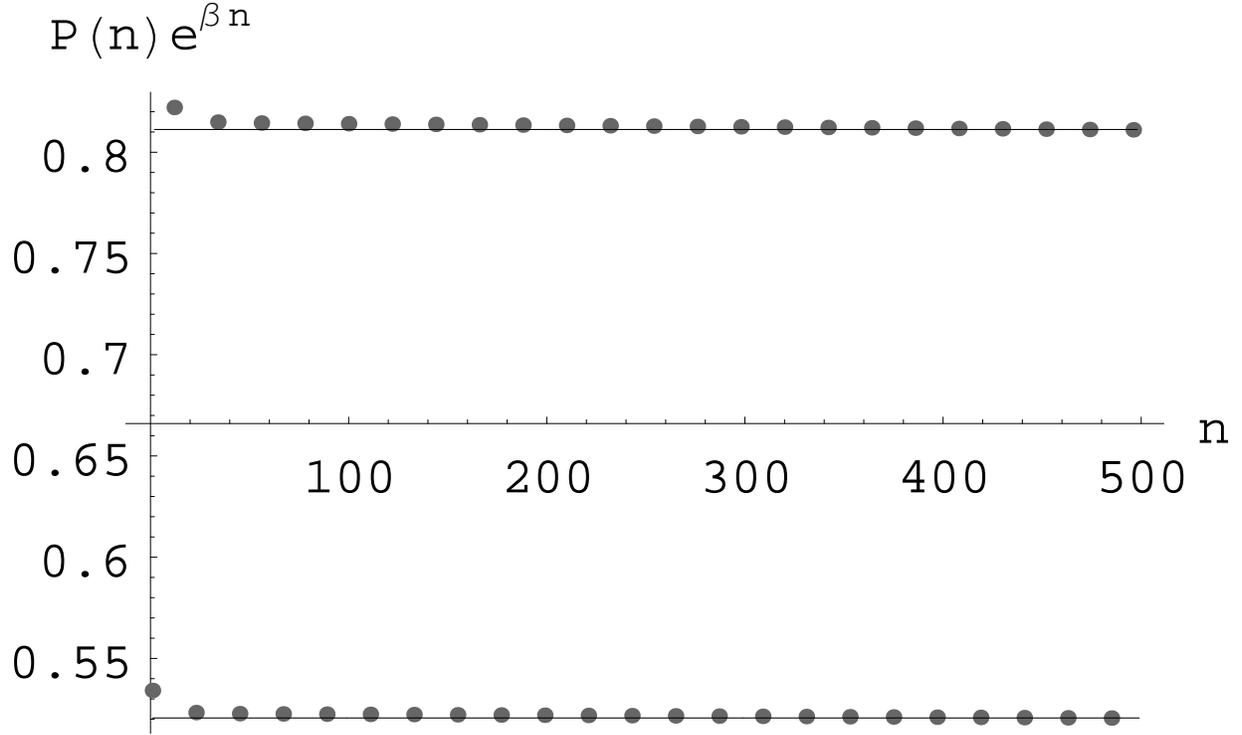}
\caption{Results of the numeric calculation of the Toeplitz
determinant are shown as $P(n) e^{\beta n}$ vs. $n$ at
$\gamma =1$, $h=1.1$. The value of $\beta$ is obtained numerically
from (\ref{betaPlus}). The solid line is the analytic result
$E(1+(-1)^n A)$ with $A=0.2182...$ from (\ref{Aplus}) and
$E=0.6659...$ obtained by fitting at large $n$. To make the plot more
readable we show only every 11th point (for $n=1,12,23,\ldots$) of the
numerical calculation of the determinant.
Note that the size of the points is not related to the estimated error
in the numerics, which is actually smaller.}
   \label{DetPlot1}
\end{figure}

Expression (\ref{oscbehavior}) for the EFP fits the numerical data 
remarkably well (see Fig.~\ref{DetPlot1}) and this fact strongly
supports the generalized Fisher-Hartwig conjecture.

One can understand these oscillations as a result of ``superconducting''
correlations of real fermions described by the Hamiltonian (\ref{realfermionH}).
Fermions are created and destroyed in pairs of nearest neighbors.
At large magnetic fields, the oscillations are due to the fact that the 
probability of having a depletion string of length $2k-1$ or $2k$ is very
similar.
Since the magnetic field in (\ref{realfermionH}) is essentially
a chemical potential for the fermions, the energy cost to destroy a pair
of particles is $4h$: at very big magnetic fields, the amplitude for a
pair destruction event is suppressed by a factor of $\gamma \over 4 h$,
i.e. a probability of $\gamma^2 \over 16 h^2$.
This means that the probability of depletion behaves like:
\bea
    P(2k-1) & \sim & 2 \left( {4 h \over \gamma} \right)^{-2k}
    \qquad {\rm and} \nonumber \\
    P(2k) & \sim & \left( {4 h \over \gamma} \right)^{-2k} \; ,
    \label{twoprob}
\eea
where the factor of two in the first expression is a simple
combinatorial factor.
The two probabilities in (\ref{twoprob}) can be combined in a
single expression:
\be
   P(n) = E \left[ 1 + A \cos(\pi n) \right]
   \left( {4 h \over \gamma} \right)^{- n} ,
   \label{Pnosc}
\ee
which is precisely (\ref{oscbehavior}), with
\be
   A = 1 - {\gamma \over h} + \Ord \left( { 1 \over h^2} \right) .
   \label{Aosc}
\ee
   
We can check the correctness of this interpretation by taking the
limit of (\ref{oscbehavior}) for $h >> 1, \gamma$.
From (\ref{betagh}) and (\ref{Aplus}) it is easy to find
\bea
   \beta(h \to \infty, \gamma) & = & \log {4 h \over \gamma}
   + \; \Ord \left( {1 \over h^2} \right) \\
   A_+ (h \to \infty, \gamma) & = & 1 - {\gamma \over h} + \Ord
   \left( {1 \over h^2} \right)
\eea
in agreement with (\ref{Pnosc},\ref{Aosc}).

%%%%%%%%%%%%%%%%%%%%%%%%%%%%%%%%%%%%%%%%
\subsection{The critical lines ($\Omega_\pm$)}
\label{critical}

%%%%%%%%%%%%%%%%%%%%%%%%
\subsubsection{$\Omega_+$ ($h = 1$)}
\label{ssOmegap}

For $h=1$ the generating function $\sigma(q)$ vanishes at $q = \pi$
and its phase has $\pi$ jumps at $q = 0, \pi$ (see
Fig.~\ref{GenFuncPlot}d).
As in the previous section, the existence of two singular points gives
rise to many terms of the form (\ref{fishgendec}).
However, in contrast to the $\Sigma_+$ region, the application of 
gFH as in (\ref{singgenexp}) shows that all terms are
suppressed by power law factors of $n$ with respect to the leading one.

The leading term is generated by the parametrization:
\be
   \sigma(q) = \tau^1(q) \eu^{{\ii \over 2} [(q - \pi)  \mod 2 \pi - \pi]}
   \left( 2 - 2 \cos (q - \pi) \right)^{1/2}
   \eu^{- {\ii \over 4} [q \mod 2 \pi - \pi]}
\ee
and consists of an exponential decay with
$\beta(1, \gamma)$ from (\ref{betaPlus}) and a power
law contribution with critical exponent $\lambda = {1\over 16}$
\be
  \left| \det ({\bf S_n}) \right| \sim
  {\it E^1_1}(\gamma) n^{- {1 \over 16} } \: \eu^{- \beta(1, \gamma) n}
  \label{leading}
\ee
with
\be
   {\it E^1_1} (\gamma) \equiv E[\tau]
   G \left( {3 \over 4} \right) G \left( {5 \over 4} \right)
   {\tau_-^{1/4} (0) \over 2^{1/4} \tau_+^{1/4} (0) \tau_- (\pi) },
\ee
where $G$ is the Barnes G-function defined in (\ref{BGfun})
and $E [\tau]$ is defined as in (\ref{Etgh}) with
\be
   \hat{\tau}_k = \hat{\sigma}_k
   + \left( {1 \over 4} - (-1)^k \right) {1 \over k} \; \theta (k)
   - {1 \over 4k} \; \theta (-k) ,
\ee
with $\hat{\sigma}_k$ from (\ref{sigmahat}).

The next term (subleading at $n\to \infty$) is obtained from
the parametrization
\be
   \sigma(q) = \tau^2(q) \eu^{-{\ii \over 2} [(q - \pi)  \mod 2 \pi - \pi]}
   \left( 2 - 2 \cos (q - \pi) \right)^{1/2}
   \eu^{\ii {3 \over 4} [q \mod 2 \pi - \pi]}
\ee
and is given by
\be
  {\it E^2_1}(\gamma) (-1)^n n^{- {9 \over 16} } \: \eu^{- \beta(1, \gamma) n}
  \label{subleading}
\ee
with
\be
   {\it E^2_1}(\gamma) \equiv E[\tau]
   G \left( {1 \over 4} \right) G \left( {7 \over 4} \right)
   {\tau_+^{3/4} (0) \over 2^{3/4} \tau_-^{3/4} (0) \tau_+ (\pi) } .
\ee

Although the inclusion of the latter (subleading) term is somewhat beyond
even gFH, we write the sum of these two terms as a
conjecture for EFP at $h=1$
\be
   P(n) \sim {\it E^1_1}(\gamma) \:
   n^{- {1 \over 16} } \left[ 1 + (-1)^n A_1(\gamma)/n^{1 \over 2}
   \right] \: \eu^{- \beta(1, \gamma) n}.
 \label{pnomp}
\ee

\begin{figure}
  \includegraphics[width=\columnwidth]{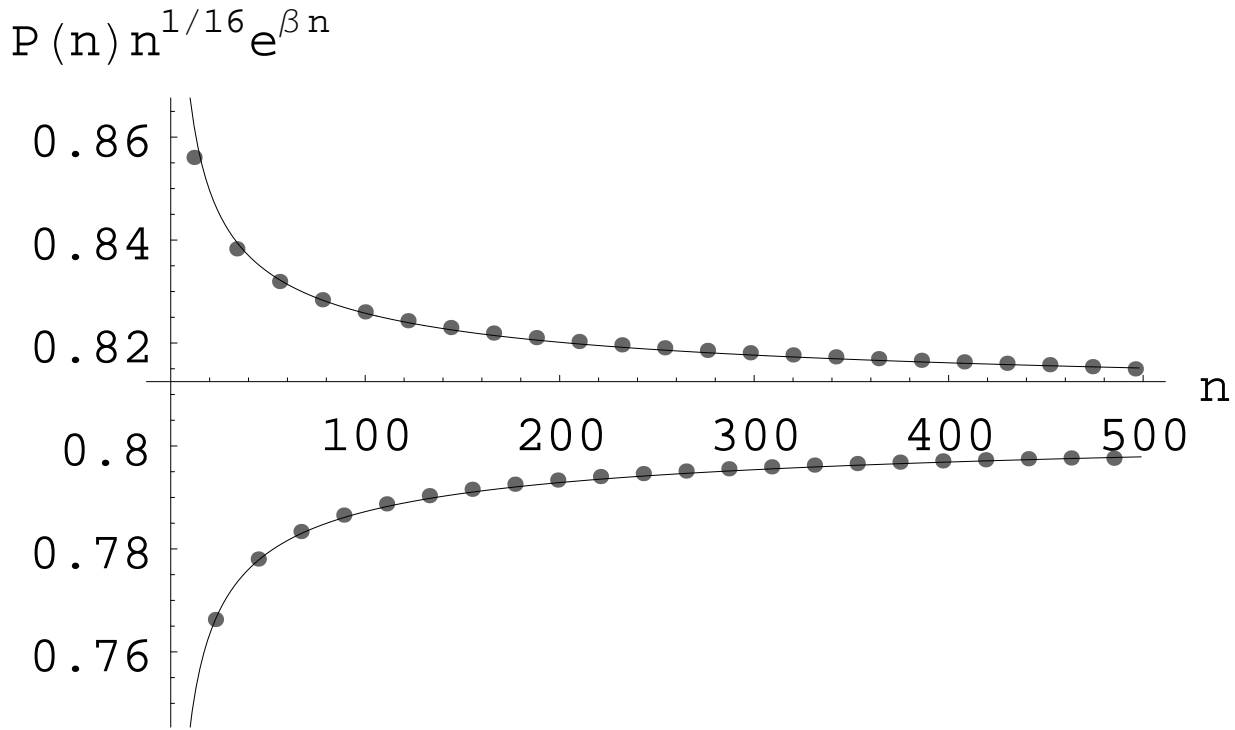}

\caption{Results of the numeric calculation of the Toeplitz determinant
are shown as $P(n) e^{\beta n} n^{1/16}$ vs. $n$ at $\gamma =1$, $h=1$.
The value $\beta=\log 2 +2G/\pi$ with Catalan's constant $G$ is
obtained from (\ref{betagh}). The solid line is the analytic result
$E(1+(-1)^n A/n^{1 \over 2})$ with $A=0.2399...$ from (\ref{A1}) and
$E=0.8065...$ as obtained by fitting at large $n$. To make the plot
more readable we show only every 11th point (for $n=1,12,23,\ldots$)
of the numerical results on the determinant.
Note that the size of the points is not related to the estimated error
in the numerics, which is actually smaller.}
   \label{DetPlot2}
\end{figure}

As these results rely on our unproven conjecture, we present our
numerical data for this case in Fig.~\ref{DetPlot2}.
Indeed, we see that the form (\ref{pnomp}) is in good agreement with
numerics and this supports our hypothesis.

The amplitude of the oscillations is
\bea
   A_1 (\gamma) & \equiv & {1 \over \sqrt{2} } \;
   { G \left( {1 \over 4} \right) G \left( {7 \over 4} \right) \over
     G \left( {3 \over 4} \right) G \left( {5 \over 4} \right) } \;
   { \tau_+ (0) \tau_- (\pi) \over \tau_- (0) \tau_+ (\pi) } \nonumber \\
   & = & {1 \over \sqrt{2} } \; { \Gamma \left( {3 \over 4} \right)
   \over \Gamma \left( {1 \over 4} \right) } \;
   { \tau (0) \over \tau (\pi) }
   \left( { \tau_- (\pi) \over \tau_- (0) } \right)^2 \nonumber \\
   & = & { \Gamma \left( {3 \over 4} \right) \over
   \Gamma \left( {1 \over 4} \right) } \; {1 \over \gamma} \;
   \exp \left( 4 \lim_{\epsilon \rightarrow 0} \oint
   {\de z \over 2 \pi} {\log \tau (z) \over z^2 - (1 + \epsilon)^2} \right),
   \label{A1In}
\eea
where we used (\ref{wienerhopf}) and the identity
\be
   G(z+1) = \Gamma (z) G(z).
\ee
To calculate the integral we deform the contour of integration as in
Fig.~\ref{SigmaStruct}e and find
\be
   A_1 (\gamma) = { \Gamma \left( {3 \over 4} \right)
   \over \Gamma \left( {1 \over 4} \right) } \; {1 \over \sqrt{2 \gamma} }.
   \label{A1}
\ee

We conclude that at $h=1$ the EFP decays exponentially at $n\to\infty$
but with an additional power law pre-factor and a damped oscillatory
component.

{\it Remark.} It is curious to notice that the exponents $1/16$ and $9/16$
in (\ref{leading}) and (\ref{subleading}) remind us of the scaling
dimensions of spins $\sigma^x$ and $\sigma^y$. \footnote{See
Ref. \cite{mccoy}, where it was shown that the power laws for the
$\sigma^x$ and $\sigma^y$ correlators are $1/4$ and $9/4$
respectively.}
It looks as if the EFP operator (\ref{EFPDef}), among other things, has
inserted square roots of spins transverse to the magnetic field at the
ends of the string.

%%%%%%%%%%%%%%%%%%%%%%%%%%%%%%
\subsubsection{$\Omega_-$ ($h = -1$)}
\label{ssOmegam}

For $h=-1$ the generating function $\sigma(q)$ does not vanish but has
a phase discontinuity of $\pi$ at $q= \pi$.
We parametrize $\sigma(q)$ as
\be
   \sigma(q) = \tau^1(q) \eu^{- {\ii \over 4} [(q - \pi)  \mod 2 \pi - \pi]}
\ee
and apply FH to obtain
\be
 \label{pnft}
   P(n) \sim {\it E^1_{-1} (\gamma)} \:
   n^{- {1 \over 16} }  \: \eu^{- \beta(-1, \gamma) n}
\ee
with
\be
   {\it E^1_{-1}} (\gamma) \equiv E[\tau]
   G \left( {3 \over 4} \right) G \left( {5 \over 4} \right)
   {\tau_-^{1/4} (\pi) \over \tau_+^{1/4} (\pi)) },
\ee
where $\beta(-1, \gamma)$ and $E [\tau]$ are defined in
(\ref{betaMinus}) and (\ref{Etgh}) with
\be
   \hat{\tau}_k = \hat{\sigma}_k
   + {(-1)^k \over 4 k} \; \theta (k)
   - {(-1)^k \over 4 k} \; \theta (-k)
\ee
and $\hat{\sigma}_k$ from (\ref{sigmahat}).

We can stretch the gFH the same way as in the previous section
for $h=+1$ by considering the second parametrization
\be
   \sigma(q) = \tau^2(q) \eu^{\ii {3 \over 4} [(q - \pi)  \mod 2 \pi - \pi]}
\ee
which gives
\be
     P'(n) \sim {\it E^2_{-1} (\gamma)} \:
     n^{- {9 \over 16} }  \: \eu^{- \beta(-1, \gamma) n}
\ee
with
\be
   {\it E^2_{-1}} (\gamma) \equiv E[\tau]
   G \left( {1 \over 4} \right) G \left( {7 \over 4} \right)
   {\tau_+^{3/4} (\pi) \over \tau_-^{3/4} (\pi)) }.
\ee
Adding this subleading term to (\ref{pnft}) we obtain
\be
   P(n) \sim {\it E^1_{-1}}(\gamma) \:
   n^{- {1 \over 16} } \left[ 1 +  A_{-1}(\gamma)/n^{1 \over 2}
   \right] \: \eu^{- n \beta(-1, \gamma)}
   \label{pnomm}
\ee
with
\bea
   A_{-1} (\gamma) & \equiv &
   { G \left( {1 \over 4} \right) G \left( {7 \over 4} \right) \over
     G \left( {3 \over 4} \right) G \left( {5 \over 4} \right) } \;
   { \tau_+ (\pi) \over \tau_- (\pi) } \nonumber \\
   & = & { \Gamma \left( {3 \over 4} \right)
   \over \Gamma \left( {1 \over 4} \right) } \;
   { \tau_+ (\pi) \over \tau_- (\pi) }.
\eea
We propose (\ref{pnomm}) as an asymptotic form for EFP at $h=-1$.

%%%%%%%%%%%%%%%%%%%%%%%%%%%%%%%%%%%%%%%%%%%%%%%%%%%%%%%%%%%%%%%%%%%
\section{The line $\Gamma_E$: an exact calculation}
\label{GammaESec}
%%%%%%%%%%%%%%%%%%%%%%%%%%%%%%%%%%%%%%%%%%%%%%%%%%%%%%%%%%%%%%%%%%%

Before we conclude our analysis of the EFP with the study of the
isotropic XY model, let us check our results (\ref{pnc},\ref{betagh})
on the special line\footnote{We are grateful to Fabian Essler who
suggested us to check our results on this special line and pointed out
the reference \cite{shrock} to us.} in the phase diagram defined by
\be
   h^2 + \gamma^2 = 1 .
   \label{trajectory}
\ee
It was shown in Ref. \cite{shrock} that on this line the ground state
is a product of single spin states and is given by
\be
   \vert G \rangle = \prod_j \vert \theta, j \rangle = \prod_j
   \left[ \cos \left( {\theta \over 2} \right) | \uparrow, j \rangle +
   (-1)^j \sin \left( {\theta \over 2} \right) | \downarrow, j \rangle
   \right],
   \label{groundstate}
\ee
where $| \uparrow, j \rangle$ is an up-spin state at the lattice
site $j$, etc.
One can directly check that the state (\ref{groundstate}) is an
eigenstate of (\ref{spinham}) if the value of parameter $\theta$ is
\be
    \cos^2 \theta  =  \frac{1 - \gamma}{1 + \gamma}
\ee
and (\ref{trajectory}) is satisfied.
It is also easy to show \cite{shrock} that this state is, in fact, the
ground state of (\ref{spinham}).

The probability of formation of a ferromagnetic string in the state
(\ref{groundstate}) is obviously
\be
   P(n) = \sin^{2n} \left( {\theta \over 2} \right)
    = \left( {1 \over 2} - {1 \over 2} { h \over |h|}
    \sqrt{ 1 - \gamma \over {1 +\gamma} } \right)^n,
 \label{pnexact}
\ee
which is an exact result on the line (\ref{trajectory}).
The value of  $\beta(h,\gamma)$ which immediately follows from this
exact result is
\be
   \beta(h = \pm \sqrt{1-\gamma^2},\gamma) =
   - \log\left( {1 \over 2} \mp {1 \over 2}
   \sqrt{ 1 - \gamma \over {1 +\gamma} } \right)
   = - \Lambda(h,\gamma) ,
 \label{betaexact}
\ee 
where $\Lambda(h,\gamma)$ was defined in (\ref{Lambda}).

This is, indeed, consistent with (\ref{beta0}) since
under the condition (\ref{trajectory}) the function
$\Delta(h,\gamma)$ vanishes. 
The integral (\ref{Delta}) defining $\Delta(h,\gamma)$ vanishes for (\ref{trajectory})
because the branching points (\ref{z1}) and (\ref{z2}) collapse to the
same point and therefore the region of integration shrinks to just one
point (\ref{Delta}).
In fact, the Toeplitz matrix (\ref{Tg}) generated by (\ref{genfunc})
becomes triangular on the line (\ref{trajectory}) with diagonal matrix
element $(S_n)_{jj} = \sin^2(\theta/2)$ and the determinant of
${\bf S_n}$ is exactly (\ref{pnexact}).

From the definitions of $\beta(h,\gamma)$, we see that the decay
factor consists of two terms, which have now a clear physical
meaning.
The $\Lambda(h,\gamma)$ term is the factor we found above in
(\ref{betaexact}) and represents the contribution given by
un-entangled spins.
The remaining part accounts for the correlations between spins.
Both $\Delta(h,\gamma)$ and the correlation functions given
by (\ref{F}) and (\ref{G}) vanish on the line 
(\ref{trajectory}).

%%%%%%%%%%%%%%%%%%%%%%%%%%%%%%%%%%%%%%%%%%%%%%%%%%%%%%%%%%%%%%%%%%%%%
\section{The critical line $\Omega_0$ ($\gamma = 0$) and the Gaussian
behavior}
\label{gammazero}
%%%%%%%%%%%%%%%%%%%%%%%%%%%%%%%%%%%%%%%%%%%%%%%%%%%%%%%%%%%%%%%%%%%%%

The case $\gamma = 0$, corresponding to the Isotropic XY Model, has
been studied in Ref. \cite{shiroishi}.
For $\gamma=0$ the generating function (\ref{genfunc}) is reduced
to the one found in \cite{shiroishi}.

For $|h| < 1$, the generating function $\sigma(q)$ has a limited
support between $[-\cos^{-1} h, \cos^{-1} h]$. To find the asymptotic
behavior of the determinant of the Toeplitz matrix one can apply
Widom's Theorem \cite{widomsupp} and obtain \cite{shiroishi}
\be
   P(n) \sim
   2^{5 \over 24} \eu^{3 \zeta'(-1)} (1-h)^{- {1 \over 8}}
   n^{- {1 \over 4}} \left( {1 + h \over 2} \right)^{n^2 \over 2}.
 \label{pnsh}
\ee
We see that in this case, the EFP decays as a Gaussian with an
additional power-law pre-factor.

In a different context, the formula (\ref{pnsh}) appeared also in
\cite{dysmehta} as a probability of forming a gap in the spectrum of
unitary random matrices.
This is not unexpected, since the joint eigenvalue distribution of
unitary random matrices is known to coincide with the distribution of free
fermions in the ground state.

For $|h| > 1$, the theory is no longer critical and the ground state
is completely polarized in the $z$ direction, giving a trivial EFP
$P(n)=0$ for $h>1$ and $P(n)=1$ for $h<-1$.

%%%%%%%%%%%%%%%%%%%%%%%%%%%%%%%%%%%%%%%%%%%%%%%%%%%%%%%%%%%
\section{Crossover between Gaussian and exponential behavior: a Bosonization approach}
\label{Crossover}
%%%%%%%%%%%%%%%%%%%%%%%%%%%%%%%%%%%%%%%%%%%%%%%%%%%%%%%%%%%

In order to understand qualitatively the crossover between the Gaussian
asymptotic behavior at $\gamma = 0$ and the exponential decay for
$\gamma \ne 0$, we employ a bosonization approach similar to
the one used in \cite{abanovkor}.
In the limit $\gamma \ll 1$ we consider the continuum
limit of (\ref{spinlessham}), bosonize the fermionic fields, and 
write the Euclidean action of the theory as
${\cal S}=\int \de x \, \de \tau {\cal L}$,
where $\tau \equiv \ii t$ is the imaginary time and the Lagrangian is
\be
   {\cal L} = 2 \sin k_F \left[ \left( \partial_\mu \vartheta \right)^2
   - {\gamma \over \pi} \cos \left( \sqrt{4 \pi} \vartheta \right)
   \right] .
   \label{bosoniclag}
\ee
Here $k_F = \cos^{-1} h$ is the Fermi momentum at $\gamma = 0$.

This is a Sine-Gordon theory for the ``conjugate field''
$\vartheta(x,\tau)$, which describes the imaginary time dynamics of our
1-D system.
In terms of $\vartheta$ the density of fermions is given by 
$\rho = \partial_{\tau} \vartheta + \rho_0$, where $\rho_{0}=k_{F}/\pi$
is the density of fermions in the ground state.

In the field theory approach, the EFP (see Ref. \cite{abanovkor}) in the
limit $n \to \infty$ would be given with exponential accuracy by the
probability of an instanton $P(n) \sim e^{-{\cal S}_{0}}$, where
${\cal S}_{0}$ is the action of the instanton.
Here the instanton is the solution of the classical equations of motion
of (\ref{bosoniclag}) which corresponds to the formation of an emptiness
of length $n$ at the time $\tau=0$.
Unfortunately, the EFP instanton involves large deviations of the
density of fermions from the equilibrium density $\rho_{0}$ and is beyond
the bosonization approach as the derivation of (\ref{bosoniclag}) relies
on the linearization of the fermionic spectrum near the Fermi points.  

Following \cite{abanovkor}, we are going to slightly generalize our
problem, by considering the depletion formation probability instead of
the EFP requiring
\be
   \left. \rho \right|_{t=0, \; 0 < x <n} = \rho_{0} +
   \left. \partial_t \vartheta (x,t) \right|_{t=0, \; 0 < x <n} 
   = \rho_{0}-\bar\rho ,
   \label{boundarycond}
\ee
where $\bar{\rho}$ is some constant.
The original EFP problem corresponds to $\bar{\rho} = \rho_{0}$. 
Here, instead, we consider the probability of weak depletion, i.e.
\be
   - \left. \partial_t \vartheta (x,t) \right|_{t=0, \; 0 < x <n}
   =\bar{\rho} << \rho_{0} .
   \label{PFWFS}
\ee
We study the latter using an instanton approach to (\ref{bosoniclag})
and infer the (qualitative) behavior of the original EFP from this weak
limit.

To simplify the problem further, we assume that the instanton
configuration is completely confined to one of the wells of the Cosine
potential in (\ref{bosoniclag}) and that the field $\vartheta$ is small enough to
allow for an expansion of the Cosine:
\be
   {\cal S} \approx 2 \sin k_F\int \de x \; \de \tau \left[ \left(
   \partial_\mu \vartheta \right)^2 + 2 \gamma \; \vartheta^2 \right] .
   \label{expandedact}
\ee
In this formulation, the anisotropy parameter $\gamma^{1/2}$ plays the
role of the mass of the bosonized theory.
The probability we are looking for is given by the action ${\cal S}_{0}$ of the
classical field configuration which satisfies the Euler-Lagrange
equation (in this case a Klein-Gordon equation in two dimensions)
with the boundary condition (\ref{boundarycond})
\be
   P_{\bar{\rho}} (n) = \eu^{-{\cal S}_0} .
   \label{actionprob}
\ee

In the limit $\gamma = 0$, the theory is massless and scale invariant.
In \cite{abanovkor} it was shown that, due to the scale invariance, the
action of the instanton is quadratic in $n$.
The instanton configuration in this case is essentially a droplet of
depletion in space-time with dimensions proportional to $n$ both in the
space and time direction, in order to satisfy the boundary condition
(\ref{boundarycond}).
This result is consistent with the Gaussian asymptotic behavior
prescribed by Widom's theorem (see Sec.~\ref{gammazero}).

In the massive case, a finite correlation length $\xi \sim
\gamma^{-1/2}$ is generated and one has a crossover behavior.
For string lengths $n$ smaller than the correlation length
$\gamma^{-1/2}$, the instanton action is not sensitive to the presence
of the finite correlation length and is still quadratic in $n$
(giving a Gaussian decay for EFP).
In the asymptotic limit of string lengths greater than $\gamma^{-1/2}$,
the time dimension of a depletion droplet is of the order of $\xi$
(instead of $n$ as in the massless limit): the action is linear in $n$
and the probability has an exponential behavior.\footnote{This
picture is very similar to the one for massless theory at finite
temperature. In the latter the inverse temperature plays the role of the
correlation length \cite{abanovkor} (see \ref{singintapp}).}

\begin{figure}
   \dimen0=\textwidth
   \advance\dimen0 by -\columnsep
   \divide\dimen0 by 2
   \noindent\begin{minipage}[t]{\dimen0}
   \includegraphics[width=\columnwidth]{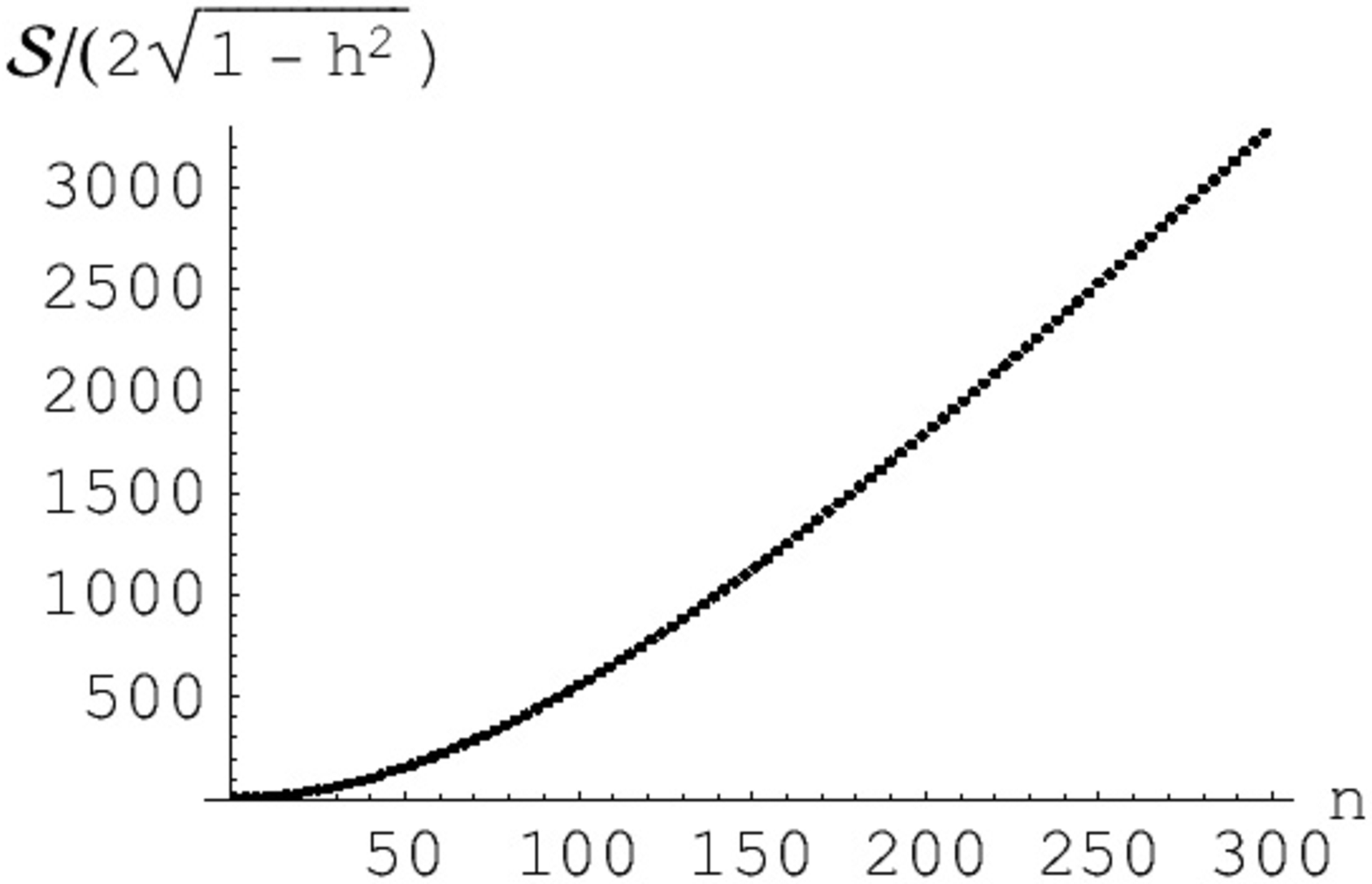}
   \caption{Plot of the value of the stationary action ${\cal S}_{0}$ vs. 
   the string length $n$. The action   ${\cal S}_{0}$
   is obtained from (\ref{statact}) with $f(y)$ given by  the
   numerical solution of the singular integral equation (\ref{singinteq}). 
   The graph depicts ${\cal S}_{0}(n)$ for $m=\sqrt{2\gamma}=0.01$, 
   $\bar{\rho}=0.2$.  The crossover takes place around $ n \sim 2/m =200$.}
   \label{actfig1}
   \end{minipage}
   \hfill
   \begin{minipage}[t]{\dimen0}
   \includegraphics[width=\columnwidth]{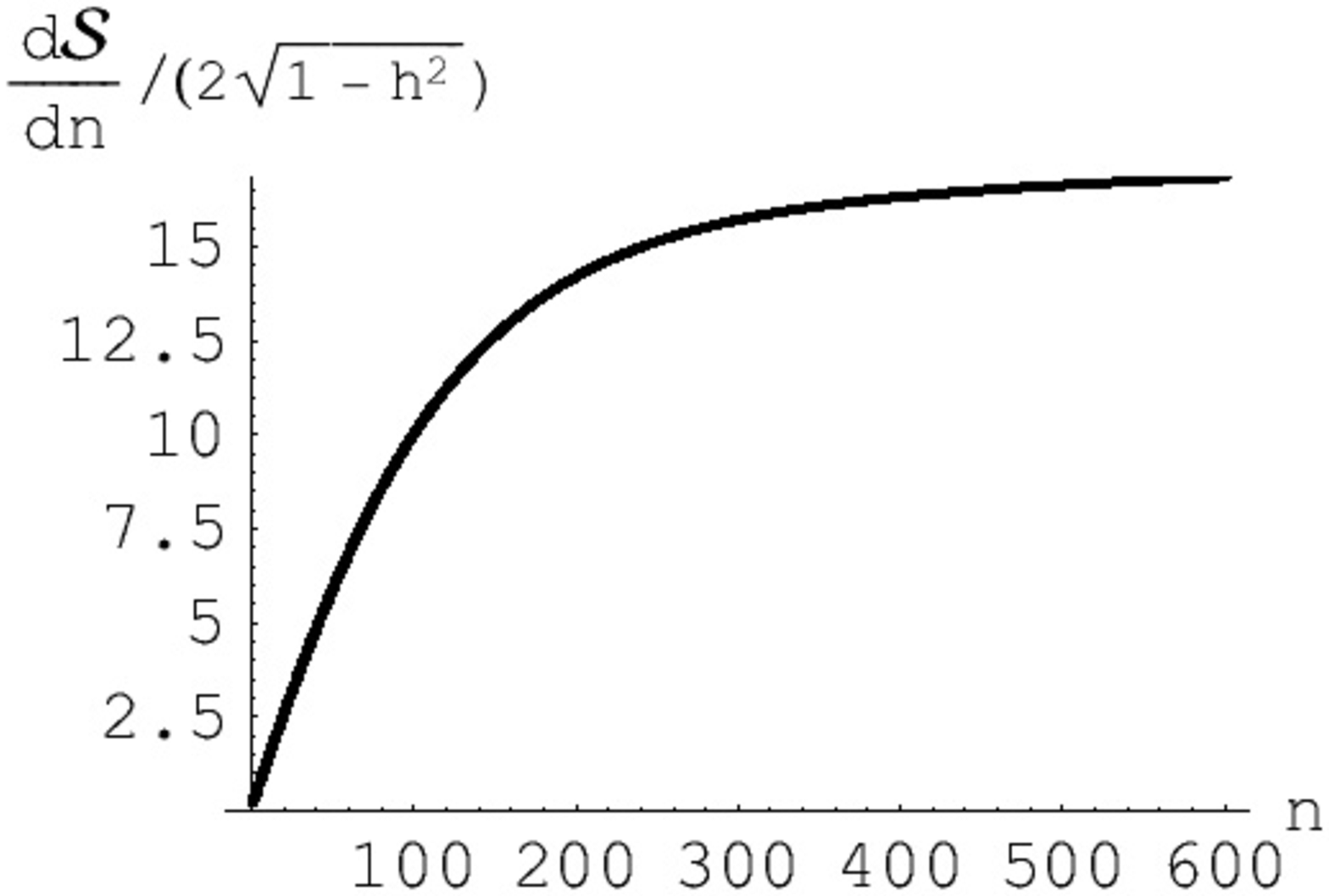}
   \caption{Plot of the derivative $d{\cal S}_{0}/dn$  with ${\cal S}_{0}$ from
   (\ref{statact}). The plot corresponds to $m=0.01$, $\bar{\rho}=0.2$ 
   and clearly shows a crossover from the quadratic to the linear behavior 
   at $n \sim 2/m = 200$.}
   \label{actfig2}
   \end{minipage}
\end{figure}

In Appendix~\ref{singintapp} we show how to solve the integral
equation corresponding to the boundary problem (\ref{boundarycond},\ref{expandedact}) and
present its numerical solution and some analytical results.
Figures \ref{actfig1} and \ref{actfig2} clearly show the
crossover between a quadratic behavior of the stationary action for
small $n$ to a linear asymptotic one for $n \to \infty$.

%%%%%%%%%%%%%%%%%%%%%%%%%%%%%%%%%%%%%%%%
\section{Discussion and Conclusions}
\label{Conclusions}
%%%%%%%%%%%%%%%%%%%%%%%%%%%%%%%%%%%%%%%%

The asymptotic behavior of the Emptiness Formation Probability $P(n)$ as
$n \to \infty$ for the Anisotropic XY model in a transverse magnetic
field as a function of the anisotropy $\gamma$ and the magnetic field
$h$ has been studied.
We have summarized our results in Table \ref{table1}.
These asymptotic behaviors have already been presented in
\cite{abanovfran}.
In this work, we completed the derivations by providing explicit
expressions for the coefficients of these asymptotic forms.

Our main motivation has been to study the relation between the
criticality of the theory and the asymptotics of the EFP.
Let us now consider the results on the critical lines ($\Omega_0$ and
$\Omega_\pm$).
The Gaussian behavior on $\Omega_0$ ($\gamma = 0$, $|h| < 1$) is in
accord with the qualitative argument of Ref. \cite{abanovkor} using a
field theory approach.
In $\Sigma_0$ ($\gamma \ne 0$, $|h| < 1$) the asymptotic decay is
exponential.
We proposed a physical interpretation of the crossover between the two
asymptotes using a bosonization analysis of the region of small
$\gamma$: we suggest that there is an intermediate regime
of Gaussian decay for the string lengths smaller than
$1 / \sqrt{\gamma}$ which crosses over to the exponential behavior for
longer strings.

On the critical lines $\Omega_\pm$, the decay of the EFP is exponential
instead of Gaussian, and apparently contradicts the qualitative picture
of Ref. \cite{abanovkor}.
The reason for this disagreement is that although at $h=\pm 1$ the
model can be rewritten in terms of massless {\it quasiparticles}
$\chi$ defined in (\ref{bogtrans}), we are still interested in the EFP
for the ``original'' Jordan-Wigner fermions $\psi$.
In terms of $\chi$ this correlator has a complicated (nonlocal)
expression very much different from the simple one (\ref{expect}).
From the technical point of view, the difference is that in the
qualitative argument in favor of a Gaussian decay of EFP for critical
systems there is an implicit assumption that the density of fermions
(or magnetization) is related in a local way to the field responsible
for the critical degrees of freedom (free boson field $\phi$).
This assumption is not valid on the lines $h=\pm 1$.
The theory is critical on those lines and can be described by some
free field $\phi$.
However, the relation between the magnetization and this field is
highly nonlocal and one can not apply the simple argument of
\cite{abanovkor} to the XY model at $h=\pm 1$.

Although EFP at the critical magnetic field does not show a Gaussian
behavior, there is an important difference between the asymptotic
behavior of EFP on and off critical lines.
Namely, a power-law pre-factor $n^{-\lambda}$ appears on all critical
lines.
For the XY model it is universal (i.e. $\lambda$ is constant on a given
critical line) and takes values $\lambda=1/4$ for $\gamma=0$
\cite{shiroishi} and $\lambda =1/16$ on the lines $h=\pm 1$.
It would be interesting to understand which operators determine
these particular ``scaling dimensions'' of the EFP (see the remark
at the end of Section \ref{ssOmegap}).

At $h\ge 1$ the use of gFH predicts even-odd oscillations of $P(n)$.
We compared the predicted oscillations to numerical calculations of
Toeplitz determinants and found a very good agreement (see
Figs. \ref{DetPlot1},\ref{DetPlot2}). 
We proposed a physical interpretation of the oscillations as coming from
pair correlations of spins which can be clearly seen as superconducting
correlations in the fermionic representation (\ref{realfermionH}).

In some parts of the phase diagram ($\Sigma_{+}$, $\Omega_{\pm}$)
we used the so-called {\it generalized Fisher-Hartwig conjecture}
\cite{basor} which is not yet proven.
However, our numeric calculations support the analytical results (see
Figures \ref{DetPlot1} and \ref{DetPlot2}).
We note that to the best of our knowledge this is the first physically
motivated example where the original Fisher-Hartwig conjecture fails
and its extended version is necessary.
\footnote{We note that recently the theory on Toeplitz determinants has
been used and extended with new results in order to calculate yet one more
important physical quantity.
We refer the interested reader to \cite{korepin03}, \cite{korepin04} and
\cite{keating}, where the entaglement for the XY Spin chain and for Random
matrix models have been calculated.
}
We also suggest that the gFH could be used to find the subleading
corrections to the asymptotic behavior, as we did for $h = \pm 1$ in
(\ref{pnomp},\ref{pnomm}).
This novel hypothesis is supported by our numerics and it would
be interesting to confirm it analytically.

In conclusion, we notice that it is straightforward to generalize our
results for nonzero temperature.
The only modification is that at $T \ne 0$ the thermal correlation
functions must be used instead of (\ref{F},\ref{G}).
Then, the generating function (\ref{genfunc}) is non-singular everywhere
and we have an exponential decay of $P(n)$ in the whole phase diagram
according to the standard Szeg\"o Theorem and standard statistical mechanics arguments.
We present results for $T \ne 0$ in Appendix~\ref{FiniteTFP}.

%%%%%%%%%%%%%%%%%%%%%%%%%%%%%%%%%%%%%%%%%%%%%%%%%%%%%%%%%%%%%
\section{Acknowledgments}
%%%%%%%%%%%%%%%%%%%%%%%%%%%%%%%%%%%%%%%%%%%%%%%%%%%%%%%%%%%%%

We greatly benefited from multiple discussions with F. Essler,
V.E. Korepin, and B.M. McCoy.
The work of AGA was supported by the NSF grant DMR-0348358, and the
Theory Institute of Strongly Correlated and Complex Systems at
Brookhaven.

%%%%%%%%%%%%%%%%%%%%%%%%%%%%%%%%%%%%%%%%
\appendix

%%%%%%%%%%%%%%%%%%%%%%%%%%%%%%%%%%%%%%%%
\section{Asymptotic behavior of Toeplitz Determinants}
\label{ToeplitzApp}

%%%%%%%%%%%%%%%%%%%%%%%%%%%%%%%%%%%%%%%%

The asymptotic behavior of the EFP for (\ref{spinham}) at
$n \to \infty$ is exactly related to the asymptotic behavior of the
determinant of the corresponding Toeplitz matrix 
(\ref{Tg},\ref{genfunc},\ref{PnX}) and can be extracted from known theorems and
conjectures in the theory of Toeplitz matrices.
These types of calculations have been done first in
\cite{LSM-1961,mccoy} for spin-spin correlation functions.
It is well known that the asymptotic behavior of the determinant of a
Toeplitz matrix as the size of the matrix goes to infinity strongly
depends upon the zeros and singularities of the generating function of
the matrix.

A very good report on the subject has been recently compiled by T. Ehrhardt
\cite{Ehrhardt-2001}.
Here we want to recap what is known about the determinant of a
Toeplitz matrix generated by a function $\sigma(q)$:
\be
   D_n [\sigma] = \det ({\bf S_n}) =
   \det \left( \int_{-\pi}^{\pi} \sigma(q)
   \eu^{- \ii (j-k) q} {\de q \over 2 \pi} \right)_{j,k=0}^n,
\ee
where the generating function $\sigma(q)$ is a periodic (complex)
function, i.e. $\sigma(q) = \sigma(2 \pi + q)$.
In this work we dealt only with generating functions with zero winding
number
\be
   {\rm Ind } \, \sigma(q) \equiv
   \int_{-\pi}^{\pi} {\de q \over 2 \pi} \, {\de \over \de q}
   \log \sigma (q) = 0
\ee
and this brief review will be limited to this condition.
This was not the case in the study of Barouch et al. \cite{mccoy},
where the generating function (see footnote after (\ref{PnX}))
had non-zero winding number in some regions of the phase
diagram.

\subsection{The Strong Szeg\"o Theorem}
%%%%%%%%%%%%%%%%%%%%%%%%%%%%%%%%%%%%%%%%

If $\sigma(q)$ is sufficiently smooth, \underline{non-zero} and
satisfies ${\rm Ind } \,\sigma(q) = 0$ (i.e., the winding number is $0$),
we can apply what is known as the {\it Strong Szeg\"o Limit Theorem}
(\cite{Hirschman}, \cite{mccoywu}), which states that the determinant
has a simple exponential asymptotic form
\be
   D_n [\sigma] \sim E[\sigma] G[\sigma]^n
   \qquad n \rightarrow \infty,
   \label{szego}
\ee
where $G[\sigma]$ and $E[\sigma]$ are defined by
\be
  G[\sigma] \equiv \exp{\hat{\sigma}_0}, \qquad
  E[\sigma] \equiv
  \exp{\sum_{k=1}^\infty k \hat{\sigma}_k \hat{\sigma}_{-k}}
  \label{szegoexp}
\ee
and $\hat{\sigma}_k$ are the Fourier coefficients of the expansion of
the logarithm of $\sigma(q)$:
\be
  \log \sigma(q) \equiv
  \sum_{k=-\infty}^\infty \hat{\sigma}_k \eu^{\ii k q}.
  \label{sigmak}
\ee

\subsection{The Fisher-Hartwig Conjecture}
%%%%%%%%%%%%%%%%%%%%%%%%%%%%%%%%%%%%%%%%

Over the years, the Szeg\"o Theorem has been extended to consider
broader classes of generating functions by relaxing the continuity
conditions which define a "smooth function", but it remained limited to
never-vanishing functions.
Therefore, some extensions have been proposed to the Szeg\"o Theorem in
order to relax this latter hypothesis.
When the generating function has only pointwise singularities (or zeros), there
exists a conjecture known as the Fisher-Hartwig Conjecture (FH)
\cite{FisherHartwig-1968}.
\footnote{This conjecture is still not completely proven.
For details and status of the conjecture see Ref. \cite{basor}.}

When $\sigma(q)$ has $R$ singularities at $q= \theta_r$ ($r=1..R$), we
decompose it as follows:
\be
   \sigma(q) = \tau(q) \prod_{r=1}^R
   \eu^{\ii \kappa_r [(q - \theta_r) \mod 2 \pi - \pi]}
   \left( 2 - 2 \cos (q - \theta_r) \right)^{\lambda_r}
   \label{fishdec}
\ee
so that $\tau(q)$ is a smooth function satisfying the conditions
stated in the previous section.
Then according to FH the asymptotic formula for the determinant takes
the form
\be
   D_n [\sigma] \sim E \left[ \tau, \{ \kappa_a \}, \{ \lambda_a \}, \{ \theta_a \} \right]
   \; n^{\sum_r \left( \lambda_r^2 - \kappa_r^2 \right)}
   G[\tau]^n \qquad n \rightarrow \infty,
   \label{singexp}
\ee
where the constant prefactor is conjectured to be
\bea
  E \left[ \tau, \{ \kappa_a \}, \{ \lambda_a \}, \{ \theta_a \} \right]
  \equiv & E[\tau] & \prod_{r=1}^R
  \tau_- \left( \eu^{\ii \theta_r} \right)^{-\kappa_r - \lambda_r}
  \tau_+ \left( \eu^{- \ii \theta_r} \right)^{\kappa_r - \lambda_r}
  \nonumber \\  & \times &
  \prod_{1 \le r \ne s \le R} \left( 1 - \eu^{\ii (\theta_s - \theta_r)}
  \right)^{(\kappa_r + \lambda_r) (\kappa_s - \lambda_s)}
  \nonumber \\ & \times &
  \prod_{r=1}^R { {\rm G} (1 + \kappa_r + \lambda_r)
  {\rm G} (1 - \kappa_r + \lambda_r) \over {\rm G} (1 + 2 \lambda_r) }.
  \label{fisherhartwig}
\eea
$E[\tau]$ and $G[\tau]$ are defined as in (\ref{szegoexp}) and
$\tau_{\pm}$ are defined by decomposition
\be
   \tau(q) = \tau_- \left( \eu^{\ii q} \right)
   G[\tau] \tau_+ \left( \eu^{- \ii q} \right),
   \label{wienerhopf}
\ee 
so that $\tau_+$ ($\tau_-$) are analytic and non-zero
inside (outside) the unit circle on which $\tau$ is defined and
satisfy the boundary conditions $\tau_+ (0) = \tau_- (\infty) = 1$.
${\rm G}$ is the {\it Barnes G-function}, an analytic entire
function defined as
\be
   {\rm G}(z + 1) \equiv (2 \pi)^{z/2} \eu^{-[z + (\gamma_E + 1) z^2]/2}
   \prod_{n=1}^\infty \left( 1 + {z \over n} \right)^k
   \eu^{-z + {z^2 \over 2n}},
   \label{BGfun}
\ee
where $\gamma_E \sim 0.57721 \ldots$ is the Euler-Mascheroni Constant.

This conjecture is actually proven  for some ranges of parameters
$\kappa_r$ and $\lambda_r$ or fully for the case of a single singularity
($R=1$), see \cite{widomsing,Ehrhardt-1997}.

In many simple cases it is possible to find  the
factorization of $\tau$ into the product of $\tau_+$ and $\tau_-$ by inspection.
More complicated examples like the ones presented in this work require
a special technique to obtain this factorization, which is known as the
{\it Wiener-Hopf decomposition}:
\bea
   \log \tau_+ (w) = \oint {\de z \over 2 \pi \ii}
   {\log \tau (z) \over z - w} & \qquad & |w| < 1, 
   \nonumber \\
   \log \tau_- (w) = - \oint {\de z \over 2 \pi \ii}
   {\log \tau (z) \over z - w} & \qquad & |w| > 1,
   \label{wienint}
\eea
where the integral is taken over the unit circle.

In light of these formulas, it is useful to present the parametrization
(\ref{fishdec}) in a form which makes the analytical structure more
apparent.
Changing the variable dependence from $q$ to $z \equiv \eu^{\ii q}$, we
write
\be
   \sigma(z) = \tau(z) \prod_{r=1}^R
   \left( 1 - {z \over z_r} \right)^{\lambda_r + \kappa_r}
   \left( 1 - {z_r \over z_{} } \right)^{\lambda_r - \kappa_r},
\ee
where $z_r \equiv \eu^{\ii \theta_r}$.

\subsection{The Generalized Fisher-Hartwig Conjecture}
\label{gfhsec}
%%%%%%%%%%%%%%%%%%%%%%%%%%%%%%%%%%%%%%%%

Despite the considerable success of the Fisher-Hartwig Conjecture, few
examples have been reported in the mathematical literature that do not
fit this result.
These examples share the characteristics that inequivalent
representations of the form (\ref{fishdec}) exist for the generating
function $\sigma(q)$.
Although no theorem has been proven concerning these cases, a
generalization of the Fisher-Hartwig Conjecture (gFH) has been suggested
by Basor and Tracy \cite{basor} that has no counter-examples yet.

If  more than one parametrization of the kind
(\ref{fishdec}) exists, we write them all as
\be
   \sigma(q) = \tau^i(q) \prod_{r=1}^R
   \eu^{\ii \kappa^i_r [(q - \theta_r) \mod 2 \pi - \pi]}
   \left( 2 - 2 \cos (q - \theta_r) \right)^{\lambda^i_r},
   \label{fishgendec}
\ee
where the index $i$ labels different parametrizations (for $R > 1$ there
can be only a countable number of different parametrizations
of this kind).
Then the asymptotic formula for the determinant is
\be
   D_n [\sigma] \sim \sum_{i \in \Upsilon}
   E \left[ \tau^i, \{ \kappa^i_a \}, \{ \lambda^i_a \}, \{ \theta_a \}
   \right] \; n^{\Omega(i)} G[\tau^i]^n \qquad n \rightarrow \infty,
   \label{singgenexp}
\ee
where
\bea
   \Omega(i) & \equiv &
   \sum_{r=1}^R \left( \left( \lambda^i_r \right)^2 -
   \left( \kappa^i_r \right)^2 \right),
 \\
   \Upsilon & = &
   \left\{ i \: \| \: \R[\Omega(i)] = \max_j \R[\Omega(j)] \right\}.
 \label{Leading}
\eea

The generalization essentially gives the asymptotics of the Toeplitz
determinant as a sum of (FH) asymptotics calculated separately for
different leading (see Eq.~(\ref{Leading})) representations
(\ref{fishgendec}). 
In Sec.~\ref{ssOmegap} we used the sum of all (not necessarily leading)
representations and showed that it also correctly produces the first
subleading corrections to the asymptotics of our Toeplitz determinant.

\subsection{Widom's Theorem}

%%%%%%%%%%%%%%%%%%%%%%%%%%%%%%%%%%%%%%%%

If $\sigma(q)$ is supported only in the interval $\alpha \le q
\le 2 \pi - \alpha$ as in our model for $\gamma=0$, singularities are
no longer pointwise and one should apply Widom's Theorem
\cite{widomsupp}.
It states that the asymptotic behavior of the determinant in this case
is
\be
   D_n [\sigma] \sim 2^{1/12} \eu^{3 \zeta'(-1)}
   \left( \sin {\alpha \over 2} \right)^{-1/4} E[\rho]^2 n^{-1/4} G[\rho]^n
   \left( \cos {\alpha \over 2} \right)^{n^2},
   \label{widomasympt}
\ee
where $E$ and $G$ are defined in (\ref{szegoexp}) and
\be
   \rho(q) = \sigma
   \left( 2 \cos^{-1} \left[\cos {\alpha \over 2} \cos q\right] \right)
\ee
with the convention $0 \le \cos^{-1} x \le \pi$.

For the case considered in Section \ref{gammazero}, the generating
function is constant, $E[\rho] = G[\rho] = 1$, and (\ref{widomasympt})
simplifies considerably giving
\be
   D_n [\sigma] \sim 2^{1/12} \eu^{3 \zeta'(-1)}
   \left( \sin {\alpha \over 2} \right)^{-1/4} n^{-1/4}
   \left( \cos {\alpha \over 2} \right)^{n^2}.
   \label{widomth}
\ee

%%%%%%%%%%%%%%%%%%%%%%%%%%%%%%%%%%%%%%%%%%%%%%%%%%%%%%%%%%%%%%%%%%%
\section{Emptiness Formation Probability at finite temperature}
\label{FiniteTFP}
%%%%%%%%%%%%%%%%%%%%%%%%%%%%%%%%%%%%%%%%%%%%%%%%%%%%%%%%%%%%%%%%%%%%

At finite temperature ($T>0$), the correlators (\ref{F}) and
(\ref{G}) become
\bea
   F_{jk}^T & \equiv & \ii \langle \psi_j \psi_k \rangle_T
   = - \ii \langle \psi_j^\dagger \psi_k^\dagger \rangle_T =
   \int_0^{2 \pi} {\de q \over 2 \pi} {1 \over 2} \sin \vartheta_q
   \tanh {\varepsilon_q \over 2 T}  \eu^{\ii q (j-k)}, 
 \\
   G_{jk}^T & \equiv & \langle \psi_j \psi_k^\dagger \rangle_T
   = \int_0^{2 \pi} {\de q \over 2 \pi} {1 \over 2}
   \left( 1+ \cos \vartheta_q  \tanh {\varepsilon_q \over 2 T} \right)
   \eu^{\ii q (j-k)}.
\eea
The EFP is expressed by (\ref{EFPNonZeroTDef}) and in the spinless
fermion formalism it becomes
\be
   P(n) = \langle \prod_{i=1}^n \psi_i \psi_i^\dagger \rangle_T.
\ee

We again use Wick's Theorem (or its thermal version, called Bloch-de
Dominicis theorem \cite{todakubo}) to express it as a Pfaffian.
The calculation proceeds the same way as for zero temperature and
the EFP can be represented as
\be
   P(n) = |\det({\bf T_n})|,
\ee
where ${\bf T_n}$ is the $n \times n$ Toeplitz matrix generated by the
function
\be
   t(q) = {1 \over 2}\left( 1 + \eu^{\ii \vartheta_q}
   \tanh {\varepsilon_q \over 2 T} \right)
\ee
where the ``rotation angle'' $\vartheta_q$ and the spectrum
$\varepsilon_q$ were defined in (\ref{rotangle}) and (\ref{spectrum})
respectively.

The generating function $t(q)$ is never-vanishing and has zero winding number.
Therefore, for $T > 0$ we can apply the standard Szeg\"o Theorem to
obtain 
\be
   P(n) {\stackrel{n \rightarrow \infty}{\sim}} E(h, \gamma, T)
   \eu^{- n \beta(h, \gamma, T)},
  \label{expbehT}
\ee
where
\bea
   \beta(h, \gamma, T) & = & -\int_0^{2 \pi} {\de q \over 2 \pi}\;
   \log \left| t(q) \right| \nonumber \\
   & = & - {1 \over 2} \int_0^{2 \pi} {\de q \over 2 \pi}\;
   \log \left[ {1 \over 2} \left( 1 + {\cos q -h \over \varepsilon_q}
   \tanh {\varepsilon_q \over 2 T} \right) \right], 
 \\
   E(h, \gamma, T) & = &  \exp \left( \sum_{k=1}^\infty k
   \hat{t}_k \hat{t}_{-k} \right)
\eea 
with 
\be
   \hat{t}_k = \int_0^{2 \pi} {\de q \over 2 \pi}
   \eu^{-\ii k q} \log \left[ {1 \over 2} \left(
   1 + {\cos q - h + \ii \gamma \sin q \over \varepsilon_q}
   \tanh { \varepsilon_q \over 2 T} \right) \right],
\ee
and $\varepsilon_q$ is given as in (\ref{spectrum}) by
\be
   \varepsilon_q = \sqrt{ (\cos q - h)^2 + \gamma^2 \sin^2 q}.
\ee

As can be expected from simple thermodynamic considerations, at finite
temperature the behavior is always purely exponential asymptotically.
As it was shown in \cite{abanovkor}, at finite but very low temperatures
one can observe a crossover from the zero temperature behavior at
short string lengths $n$ to the exponential behavior (\ref{expbehT}) in the
limit of very large $n$.
This crossover occurs at a length scale of the order of the inverse
temperature.

%%%%%%%%%%%%%%%%%%%%%%%%%%%%%%%%%%%%%%%%%%%%%%%%%%%%%%%%%%%%%%%%%%%
\section{Calculation of the stationary action in the bosonization approach}
\label{singintapp}
%%%%%%%%%%%%%%%%%%%%%%%%%%%%%%%%%%%%%%%%%%%%%%%%%%%%%%%%%%%%%%%%%%%%

In Section \ref{Crossover} we have formulated the XY
model near $\gamma = 0$ in terms of the bosonic field with Lagrangian
(\ref{expandedact}).
It was also pointed out that, instead of the EFP, we are interested in
the Probability of Formation of Weakly Ferromagnetic Strings (PFWFS)
and that we are going to calculate this probability in the saddle point
approximation.
Therefore, we consider a configuration of the field (instanton)
which satisfies the boundary condition imposed by the PFWFS
(\ref{boundarycond},\ref{PFWFS})
\be
   \left. \partial_t \vartheta (x,t) \right|_{t=0,0<x<n} = \bar{\rho}
   \label{BoundaryCond}
\ee
and that minimizes the action, i.e. that satisfies the Euler-Lagrange
equations:
\be
   \left( \partial_\mu \partial^\mu - m^2 \right) \vartheta = 0.
   \label{kleingordon}
\ee
The latter equation is the Klein-Gordon equation with the mass given by
$m^2 \equiv 2 \gamma$ (see (\ref{expandedact})).
The PFWFS will be found from the value of the action ${\cal S}_{0}$ 
corresponding to this
instanton configuration (\ref{actionprob}).
In this appendix we calculate the stationary action needed in
Sec.~\ref{Crossover}.

We now solve the differential equation (\ref{kleingordon}) with non-trivial
boundary condition (\ref{BoundaryCond}) by recasting it as the integral
equation:
\be
   \vartheta (x,t) = {1 \over 2 \pi} \int_0^n \partial_t K_0
   \left( m \sqrt{(x-y)^2 +t^2} \right) f(y) \; \de y,
\ee
where $K_0 (x,x';t,t')$ is the modified Bessel function of 0-th order
-- the kernel of the differential operator (\ref{kleingordon}) in two
dimensions.
We impose the boundary condition (\ref{BoundaryCond}) by requiring that
the ``source'' $f(y)$ satisfies
\bea
   \left. \partial_t \vartheta(x,0) \right|_{0<x<n} & = &
   \lim_{t \to 0} {1 \over 2 \pi} \int_0^n \left\{
   K_2 \left( m \sqrt{(x-y)^2 +t^2} \right) { m^2 t^2 \over (x-y)^2 
   + t^2 } \right. \nonumber \\
   && \quad \left. - \,
   K_1 \left( m \sqrt{(x-y)^2 +t^2} \right) { m \over \sqrt{(x-y)^2
   + t^2 } } \right\} f(y) \; \de y = \bar{\rho} .
   \label{inteq}
\eea
This is the integral equation on $f(y)$ we have to solve.

Once the limit $t \to 0$ is taken, the kernel in Eq. (\ref{inteq}) is
singular.
We isolate the singularity by rewriting equation (\ref{inteq}) as:
\be
   {\de \over \de x} \; \dashint_0^n {f(y) \over x - y} \; \de y
   + \lim_{t \to 0} \int_0^n G_0 (x,t;y) f(y)\; \de y = 2 \pi \bar{\rho}
\ee
with
\bea
   G_0 (x,t;y) & \equiv & {(x-y)^2 - t^2 \over (x-y)^2 + t^2} +
   K_2 \left( m \sqrt{(x-y)^2 +t^2} \right) { m^2 t^2 \over (x-y)^2 +t^2 }
   \nonumber \\
   && -
   K_1 \left( m \sqrt{(x-y)^2 +t^2} \right) { m \over \sqrt{(x-y)^2 +t^2 } }
   \label{kernel0}
\eea
or, after integration over $x$, as 
\be
   \dashint_0^n {f(y) \over x - y} \; \de y
   + \int_0^n G (x;y) f(y)\; \de y = 2 \pi \bar{\rho} \; x
   \label{singinteq}
\ee
with
\be
   G(x;y) \equiv \lim_{t \to 0} \int_0^x G_0 (x_1,t;y) \; \de x_1 .
\ee
We have recasted Eq.~(\ref{inteq}) in the standard form for a
singular integral equation (\ref{singinteq}).
Once we have the solution of this equation, we can calculate the 
action corresponding to this instanton as
\be
   {\cal S}_0 = 4 \sqrt{1 - h^2} \bar{\rho} \int_0^n f(y) \; \de y.
   \label{statact}
\ee

We solved singular integral equation (\ref{singinteq}) numerically and
we computed the corresponding action (\ref{statact}).
The results of these calculation are presented as a plot of the action
${\cal S}_{0}$ vs. $n$ in Fig.~\ref{actfig1}, where we notice the 
crossover from a quadratic to a linear behavior (corresponding to a
crossover from Gaussian to exponential behavior for the probability, 
(\ref{actionprob})) as we expected.
To confirm the nature of this crossover, in Fig.~\ref{actfig2} we 
plot $d{\cal S}_{0}/dn$ and we see that it starts
linearly and then saturates asymptotically as it should.

In the limit $n \ll 1/m$, we can expand the Bessel functions in the kernel
(\ref{kernel0})
\be
   G_0 (x,t;y) =  -{m^2 \over 2} \left( {t^2 \over (x-y)^2 +t^2} +
   {1 \over 2} \ln \left[ (x-y)^2 + t^2 \right]
   + \ln {m \over 2} + G - {1 \over 2} \right) + \ldots,
   \label{G0exp}
\ee
where $G$ is Catalan's constant.
Then we solve the singular integral equation (\ref{singinteq}) to
first order by first transforming it into a regular integral equation.

In \cite{musk}, Chap. 14, Sec. 114 it is explained that a singular
integral equation like (\ref{singinteq}) is equivalent to
\be
   f(x) + {1 \over \pi \ii} \int_0^n N(x;y) \; f(y) \; \de y
   = 2 \bar{\rho} \sqrt{x(n-x)},
   \label{reginteq}
\ee
where the new kernel is
\be
   N(x;y) \equiv {\sqrt{x(n-x)} \over \pi \ii} \; \dashint_0^n
   {G(y';y) \over \sqrt{y'(n-y')} (x-y')} \; \de y' .
\ee

\begin{figure}
   \includegraphics[width=\columnwidth]{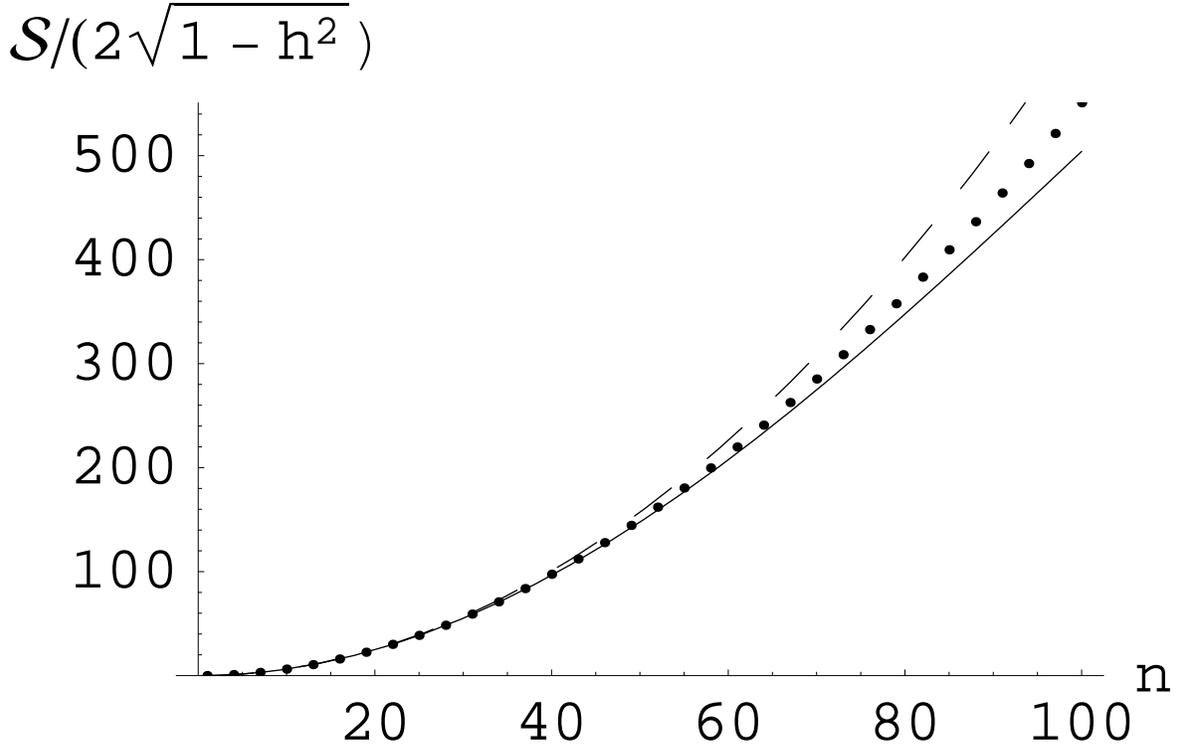}
   \caption{The solid line is the plot of the stationary action
(\ref{ansol}) against $n$. This analytical solution is valid for
$n << 1/m$ and corresponds to $m = 0.01$ and $\bar{\rho} = 0.2$.
The dotted line represents the value of the action (\ref{statact})
with the source given by numerical solution of the singular integral
equation (\ref{singinteq}).
The dashed line corresponds to the zeroth-order, pure Gaussian, solution,
i.e. (\ref{ansol}) with $m \equiv 0$, which we include for comparison.
We see that the inclusion of first order correction almost doubles the
range in which the analytical solution is accurate.}
   \label{actcomp}
\end{figure}

Using (\ref{G0exp}), we can explicitly calculate the integral defining
$N(x;y)$ in terms of elementary functions and after some algebra the
integral equation (\ref{reginteq}) results in a long, but essentially
simple, regular integral equation.
Its solution is
\be
   f(x) = \bar{\rho} \left[ 2 + {m^2 n^2 \over 8} \left(
   \ln {m \, n \over 8} + G - {3 \over 2} \right) \right] \sqrt{x(n-x)}
   - \bar{\rho} \; {m^2 n^2 \over 4} \left(x - {n \over 2} \right)
   \tan^{-1} \sqrt{ x \over n - x}.
\ee
The corresponding stationary action (\ref{statact}) is
\be
   {\cal S}_0 = \pi \sqrt{1 - h^2} \bar{\rho}^2 n^2 \left[ 1 +
   {m^2 n^2 \over 16} \left( \ln {m \; n \over 8} + G - 2 \right) \right] .
   \label{ansol}
\ee
The first term in (\ref{ansol}) corresponds to the Gaussian decay of
PFWFS we expect in the limit of $m=0$.
In Fig.~\ref{actcomp}, we compare this analytical result for the action
with the numerical result of Fig.\ref{actfig1}.
In the plot, we include the pure Gaussian decay (the first term in
(\ref{ansol})), which already gives a remarkable agreement for small $n$.
The full solution (\ref{ansol}) extends this agreement further for larger
$n$.

%%%%%%%%%%%%%%%%%%%%%%%%%%%%%%%%%%%%%%%%

\end{document}